\documentclass[12pt]{iopart}
\usepackage{graphicx}
\usepackage{dcolumn}
\usepackage{bm}

\newcommand{\be}{\begin{equation}}
\newcommand{\ee}{\end{equation}}
\newcommand{\ba}{\begin{array}}
\newcommand{\ea}{\end{array}}
\newcommand{\bn}{\begin{eqnarray}}
\newcommand{\en}{\end{eqnarray}}
\newcommand{\bnl}{\begin{mathletters}\begin{eqnarray}}
\newcommand{\enl}{\end{eqnarray}\end{mathletters}}
\newcommand{\bml}{\begin{mathletters}}
\newcommand{\eml}{\end{mathletters}}
\newcommand{\bc}{\begin{center}}
\newcommand{\ec}{\end{center}}
\newcommand{\bt}{\begin{tabular}}
\newcommand{\et}{\end{tabular}}
\newcommand{\bnll}[1]{\begin{eqnarray}}
\newcommand{\enll}{\end{eqnarray}}
\newcommand{\bw}{}
\newcommand{\ew}{}

\newcommand{\half}{\frac{1}{2}}
\newcommand{\tfrac}[2]{{\textstyle{\frac{#1}{#2}}}}
\newcommand{\thalf}{\tfrac{1}{2}}

\renewcommand{\rmd}{{\rm d}}

\newcommand{\pr}{\phantom{-}}
\newcommand{\prac}{        {\frac{4}{5}}}
\newcommand{\phac}{\phantom{\frac{4}{5}}}

\newcommand{\bbox}[1]{\bm{#1}}
\newcommand{\scalr}{{r}}
\newcommand{\bboxr}{\bbox{r}}
\newcommand{\bboxR}{\bbox{R}}
\newcommand{\bboxrp}{\bbox{r}'}

\newcommand{\text}[1]{\mathrm{#1}}

\font\bigint=cmex10 scaled 1600

\newcommand{\hint}{{\bigint\raisebox{0.7em}{\char'122}}}

\begin{document}

\title{The Negele-Vautherin density matrix expansion
applied to the Gogny force}

\author{J. Dobaczewski}
\address{Institute of Theoretical Physics, University of Warsaw,
             Ho\.za 69, PL-00681 Warsaw, Poland \\
         Department of Physics, P.O. Box 35 (YFL),
             FI-40014 University of Jyv\"askyl\"a, Finland}

\author{B.G. Carlsson}
\address{ Department of Physics, P.O. Box 35 (YFL),
             FI-40014 University of Jyv\"askyl\"a, Finland}

\author{M. Kortelainen}
\address{Department of Physics \&
  Astronomy, University of Tennessee, Knoxville, Tennessee 37996, USA \\
         Physics Division, Oak Ridge National Laboratory, P.O. Box
  2008, Oak Ridge, Tennessee 37831, USA}

\date{, 53th draft: February 9, 2010, today: \today}

\begin{abstract}

We use the Negele-Vautherin density matrix expansion to derive a quasi-local
density functional for the description of systems of fermions interacting with
short-ranged interactions composed of arbitrary
finite-range central, spin-orbit, and tensor components. Terms
that are absent in the original Negele-Vautherin approach owing to the
angle averaging of the density matrix are fixed by employing a
gauge invariance condition.
We obtain the Kohn-Sham interaction energies in all spin-isospin
channels, including the exchange terms, expressed as functions of the
local densities and their derivatives up to second (next to leading)
order. We illustrate the method by determining the coupling constants
of the Skyrme functional or Skyrme force that correspond to the
finite-range Gogny central force. The resulting self-consistent
solutions reproduce the Gogny-force binding energies and radii within
the precision of 1-2\%.

\end{abstract}


\section{Introduction}\label{sec1}

The search for a universal energy density functional (EDF)
\cite{[Ber07a]}, that would be able to provide a spectroscopic-quality
\cite{[Zal08]} description of atomic nuclei, is at the focus of the
present-day studies in nuclear structure. Recently, several advanced
phenomenological analyses were aimed at improving the standard
relativistic \cite{[Lon08]} or nonrelativistic local
\cite{[Les07],[Zal08],[Car08],[Ben09b],[Zal10]} or nonlocal functionals \cite{[Cha08]}. A
significant ongoing effort is also devoted to microscopic
derivations of the functionals (see, for example,
Refs.~\cite{[Fin06],[Bog09],[Geb09]}.) One of the central points of
the current EDF studies is the question: to what extent can the finite-range
effective interactions be approximated as quasi-local density
functionals?

The framework to build the quasi-local theory was set up in the seminal paper by
Negele and Vautherin (NV) \cite{[Neg72]}, which introduced the so-called
density matrix expansion (DME) method. Later, other methods to achieve the
same goal, like the semiclassical expansion \cite{[Sou00],[Sou03]}
were also proposed and studied. The original NV expansion, for the scalar density,
allowed for treating only the even-order terms in relative coordinates, and
thus was applicable only to even-power gradient densities. In the present study,
we revisit the NV expansion by adding the odd-power gradient densities
through the local gauge-invariance condition. There are two other important
differences of the present approach with respect to the NV original,
namely, (i) we treat the spin (vector) densities analogously to the scalar
densities and (ii) we apply a different DME for the direct terms. These differences
are motivated by the effective-theory \cite{[Lep97]} interpretation of the DME
advocated in the present study. The ultimate test of these ideas
can only be obtained by analysing microscopic properties of
nuclear densities \cite{[Geb09]}.

We present a complete
set of expressions in all spin-isospin channels, applicable to
arbitrary finite-range central, spin-orbit, and tensor interactions.
Our study is focused on applying the NV expansion to the Gogny
interaction~\cite{[Dec80],[Ber91b]}. This allows us to look at the
correct scale of the interaction range that properly characterizes
nuclear low-energy phenomena. By performing expansion up to second
order (or next-to-leading order, NLO), one obtains the local Skyrme
functional \cite{[Ben03],[Per04]}. In this way, we establish
a firm link between two different, local and nonlocal, very successful
functionals.

The paper is organized as follows. In Section \ref{sec2}, we introduce
reformulation of the NV expansion in terms of particles without
spin. Discussion of this example allows us to present details of the
approach without complications that otherwise could have obscured the main
ideas. In Sections~\ref{sec3} and \ref{sec4}, we present results with
spin and isospin degrees of freedom reintroduced, with
Section~\ref{sec4} containing applications of the formalism to the
Gogny force. Conclusions are given in Section \ref{sec9}, and
\ref{appA} contains the discussion of spin and isospin polarized
nuclear matter. Preliminary results of the present study were
published in Ref.~\cite{[Dob03b]}.

\section{Local energy density for spinless particles of one kind}\label{sec2}

In this section, we consider the simplest (and academic) case of fermions
with no spin and no isospin.
First we recall that for an arbitrary non-local finite-range interaction
$V(\bboxrp_1,\bboxrp_2;\bboxr_1,\bboxr_2)$,
the Kohn-Sham interaction energy \cite{[Koh65a]} has the form
\bn
   {\cal E}^{\text{int}}
    &=& \thalf\int\rmd^3\bboxrp_1\rmd^3\bboxrp_2\rmd^3\bboxr_1\rmd^3\bboxr_2
            V(\bboxrp_1, \bboxrp_2; \bboxr_1, \bboxr_2)
      \times  \nonumber \\ && ~~~~~~~~~~~~
        \big(\rho(\bboxr_1,\bboxrp_1)\rho(\bboxr_2,\bboxrp_2)
      -      \rho(\bboxr_2,\bboxrp_1)\rho(\bboxr_1,\bboxrp_2)\big) ,
\en
whereas for a local interaction,
   \be
   V(\bboxrp_1,\bboxrp_2;\bboxr_1,\bboxr_2) =
     \delta(\bboxrp_1\!-\!\bboxr_1)\delta(\bboxrp_2\!-\!\bboxr_2)
   V(\bboxr_1,\bboxr_2) ,
   \ee
the interaction energy reduces to:
   \bn
   {\cal E}^{\text{int}}
    &=& \thalf\int\rmd^3\bboxr_1\rmd^3\bboxr_2
            V(\bboxr_1, \bboxr_2)
\label{eq01}
        \Big(\rho(\bboxr_1)\rho(\bboxr_2)
      -      \rho(\bboxr_2,\bboxr_1)\rho(\bboxr_1,\bboxr_2)\Big),
   \en
where $\rho(\bboxr_1)\equiv\rho(\bboxr_1,\bboxr_1)$
and   $\rho(\bboxr_2)\equiv\rho(\bboxr_2,\bboxr_2)$ are local densities.
As is well known, the first term in Eq.~(\ref{eq01}) (the direct term)
depends only on local densities, whereas the second one (the exchange term)
depends on the modulus squared of the non-local density. This markedly
different structure of the two terms requires separate treatment,
as discussed in the following two subsections.

\subsection{Direct interaction energy}\label{sec2a}

In nuclei, the range of
interaction is significantly smaller than the typical scale of
the distance at which the local density varies. Therefore, we
may expand the local densities $\rho(\bboxr_1)$ and $\rho(\bboxr_2)$
around their average position, and use this
expansion to calculate the direct term in Eq.~(\ref{eq01}).

Denoting the standard total ($\bboxR$) and relative
($\bboxr$) coordinates and derivatives as
   \be\label{eq04}
   \bboxR  = \thalf(\bboxr_1 + \bboxr_2) , \quad\quad
   \bboxr  =        \bboxr_1 - \bboxr_2  ,
   \ee
   \be
\label{eq05b}
   \bbox{\nabla}  = \frac{\partial}{\partial\bboxR}
                  = \frac{\partial}{\partial\bboxr_1}
                  + \frac{\partial}{\partial\bboxr_2} , \quad\quad
   \bbox{\partial}= \frac{\partial}{\partial\bboxr}
                  = \half\left(\frac{\partial}{\partial\bboxr_1}
                  -            \frac{\partial}{\partial\bboxr_2}\right) ,
   \ee
we have the expansion of local densities to second order,
\bw
   \bn
   \label{eq03a}
   \rho(\bboxr_1) = \rho(\bboxR + \thalf\bboxr) &=&
                    \rho(\bboxR)+ \thalf r_a\nabla_a\rho(\bboxR)
                                + \tfrac{1}{8} r_a r_b
                                  \nabla_a\nabla_b\rho(\bboxR) + \ldots , \\
   \label{eq03b}
   \rho(\bboxr_2) = \rho(\bboxR - \thalf\bboxr) &=&
                    \rho(\bboxR)- \thalf r_a\nabla_a\rho(\bboxR)
                                + \tfrac{1}{8} r_a r_b
                                  \nabla_a\nabla_b\rho(\bboxR) + \ldots ,
   \en
and hence
   \be
\label{eq17}
   \rho(\bboxr_1)\rho(\bboxr_2) = \rho^2(\bboxR)
    +\tfrac{1}{4}r_ar_b
                           \Big(\rho(\bboxR)\nabla_a\nabla_b\rho(\bboxR)
              -[\nabla_a\rho(\bboxR)][\nabla_b\rho(\bboxR)]\Big) + \ldots,
   \ee
\ew
where we implicitly assumed the summation over the repeated Cartesian
indices $a$ and $b$.

Assuming that the local potential $V(\bboxr_1,\bboxr_2)$ depends only
on the distance between the interacting particles,
$V(\bboxr_1,\bboxr_2)$=$V(|\bboxr_1$$-$$\bboxr_2|)$=$V(r)$, the direct
interaction energy is given by the integral of a local energy
density ${\cal H}^{\text{int}}_{\text{dir}}(\bboxR)$,
   \be
\label{eq19}
   {\cal E}^{\text{int}}_{\text{dir}}
    = \int\rmd^3\bboxR \,{\cal H}^{\text{int}}_{\text{dir}}(\bboxR),
   \ee
where up to second order,
   \be
\label{eq19a}
    {\cal H}^{\text{int}}_{\text{dir}}(\bboxR)
    = \thalf\Big[ V_0 \rho^2 + \tfrac{1}{12} V_2
                           \Big(\rho\Delta\rho
              -(\bbox{\nabla}\rho)^2\Big)\Big] + \ldots,
   \ee
where the coupling constants, $V_0$ and $V_2$, are given by the lowest
two moments of the interaction,
   \be\label{eq02}
  V_n = \int\rmd^3\bboxr\,r^n V(r)
                           = 4\pi \int\rmd{r}\,r^{n+2} V(r) .
   \ee
After integrating by parts, the two second-order terms in Eq.~(\ref{eq19a})
are identical, so we can equally well use:
   \be
\label{eq19b}
    {\cal H}^{\text{int}}_{\text{dir}}(\bboxR)
    = \thalf\Big[ V_0 \rho^2 + \tfrac{1}{6} V_2
                           \rho\Delta\rho
              \Big] + \ldots,
   \ee

We see that the separation of scales between the range of
interaction and the rate of change of the local density leads to a
dramatic collapse of information that is transferred from the
interaction potential to the interaction energy. Namely, the two
constants, $V_0$ and $V_2$, completely characterize the interaction
in the direct term, and the detailed form
of the potential $V(r)$ becomes irrelevant. Moreover, it can be easily checked
that in this
approximation, the direct energy density is exactly equal to that
corresponding to the contact force
corrected by the second-order gradient pseudopotential,
\be
\tilde{V}(\bboxr) = V_0\delta(\bboxr)-\tfrac{2}{3}V_2\bbox{\partial}
                  \cdot\delta(\bboxr)\bbox{\partial} .
\ee

\subsection{Exchange interaction energy}\label{sec2b}

In the exchange term of Eq.~(\ref{eq01}), the non-zero range of the interaction
probes the non-local space dependence of the density matrix.
For short-range interactions, one can expand
$\rho(\bboxR,\bboxr)$ to second order with respect to the variable $\bboxr$, which
gives
\bw
\bn\!\!\!\!\!\!\!\!\!\!\!\!\!\!\!\!
   \rho(\bboxr_1,\bboxr_2)  = \rho(\bboxR,\bboxr) &=&
                    \rho(\bboxR)
                  + r_a\partial_a\rho(\bboxR,\bboxr)
                  +\thalf r_a r_b
                         \partial_a\partial_b\rho(\bboxR,\bboxr) + \ldots
\nonumber \\
\label{eq08}
                           &=&
                    \rho(\bboxR)
                  + i r_aj_a(\bboxR)
                  + \tfrac{1}{2} r_a r_b
           \left[\tfrac{1}{4}\nabla_a\nabla_b\rho(\bboxR)-\tau_{ab}(\bboxR)\right] + \ldots ,
\en
where derivatives $\partial_i$ are always
calculated at $r_a$=0, and therefore, the result can be expressed in
terms of the standard current and kinetic densities \cite{[Eng75]}:
\be
\label{eq38}
   j_a(\bboxR)   =\tfrac{1}{i}\partial_a\rho(\bboxR,\bboxr)_{\bboxr=0}, \quad\quad
\tau_{ab}(\bboxR)=\nabla_a^{(1)}\nabla_b^{(2)}\rho(\bboxr_1,\bboxr_2)_{\bboxr_1=\bboxr_2} .
\ee

This parabolic approximation does not ensure that
$\rho(\bboxr_1,\bboxr_2)$$\longrightarrow$0
for large $\scalr=|\bboxr|=|\bboxr_1$$-$$\bboxr_2|$.
In the spirit of the DME \cite{[Neg72]}, one can improve it
by introducing three functions of $\scalr$,
$\pi_0(\scalr)$, $\pi_1(\scalr)$, and $\pi_2(\scalr)$ \cite{[Dob03b]}
that vanish at large $\scalr$, i.e.,
we define the quasi-local approximation of the density matrix by:
   \bn\!\!\!\!\!\!\!\!\!\!\!\!\!\!\!\!\!\!\!\!\!\!\!\!\!\!\!\!\!\!\!\!
   \rho(\bboxr_1,\bboxr_2)   &\!\!\!\!\!\!\!\!\!\!=&
                     \pi_0(\scalr)\rho(\bboxR)
                    +  \pi_1(\scalr)  r_a\partial_a\rho(\bboxR,\bboxr)
              + \thalf \pi_2(\scalr) r_a r_b
                         \partial_a\partial_b\rho(\bboxR,\bboxr) + \ldots
\nonumber \\
                             &\!\!\!\!\!\!\!\!\!\!=&
                     \pi_0(\scalr)\rho(\bboxR)
                    +  i\pi_1(\scalr)  r_aj_a\rho(\bboxR)
              + \tfrac{1}{2} \pi_2(\scalr) r_a r_b
\left[\tfrac{1}{4}\nabla_a\nabla_b\rho(\bboxR)-\tau_{ab}(\bboxR)\right] + \ldots~.
\label{eq09}
   \en
Such a postulate has to be compatible with the Taylor
expansion of Eq.~(\ref{eq08}), which requires that
\be
\pi_0(0)=\pi_1(0)=\pi_2(0)=1 \quad\mbox{and}\quad
\pi'_0(0)=\pi'_1(0)=\pi''_0(0)=0.
\ee
Of course, for
$\pi_0(\scalr)$=$\pi_1(\scalr)$=$\pi_2(\scalr)$=1, one reverts
to the parabolic approximation (\ref{eq08}).

The product of nonlocal densities in the exchange integral
of Eq.~(\ref{eq01}) to second order reads
   \bn\!\!\!\!\!\!\!\!\!\!\!\!\!\!\!\!\!\!\!\!\!\!\!\!\!\!\!\!\!\!\!\!
   \rho(\bboxr_1,\bboxr_2)\rho(\bboxr_2,\bboxr_1) &=&
                 \pi_0^2(\scalr)\rho^2(\bboxR)
\nonumber \\
              &&\!\!\!\!\!\!\!\!\!\!\!\!\!\!\!\!\!\!\!\!\!\!\!\!\!\!\!\!\!\!\!\!
               + \pi_0(\scalr)\pi_2(\scalr)r_a r_b\Big\{
                \rho(\bboxR)\partial_a\partial_b\rho(\bboxR,\bboxr)
               -              [\partial_a\rho(\bboxR,\bboxr)]
                              [\partial_b\rho(\bboxR,\bboxr)]\Big\} + \ldots
\nonumber \\
              &=&\pi_0^2(\scalr)\rho^2(\bboxR)
\nonumber \\
\label{eq10}
              &&\!\!\!\!\!\!\!\!\!\!\!\!\!\!\!\!\!\!\!\!\!\!\!\!\!\!\!\!\!\!\!\!
               + \pi_0(\scalr)\pi_2(\scalr) r_a r_b
\Big\{\tfrac{1}{4}\rho(\bboxR)\nabla_a\nabla_b\rho(\bboxR)-\rho(\bboxR)\tau_{ab}(\bboxR)
              +j_a(\bboxR)j_b(\bboxR)\Big\} + \ldots~,
   \en
\ew
where we have introduced the supplementary condition:
\be
\pi_1^2(\scalr)=\pi_0(\scalr)\pi_2(\scalr).
\label{eq40}
\ee
This condition ensures that the quasi-local approximation of Eq.~(\ref{eq09}) is compatible
with the local gauge invariance \cite{[Dob95e]}.
Indeed, the left-hand side of Eq.~(\ref{eq10}) is manifestly invariant
with respect to the local gauge transformation,
\be
\rho'(\bboxr_1,\bboxr_2) = e^{i\phi(\bboxr_1)-i\phi(\bboxr_2)}
\rho (\bboxr_1,\bboxr_2) ,
\ee
and only the difference of terms in the curly brackets in Eq.~(\ref{eq10})
is invariant with respect to the same transformation \cite{[Eng75],[Per04]}.

Functions $\pi_0(\scalr)$, $\pi_1(\scalr)$, and $\pi_2(\scalr)$
also depend on the parameters defining the approximation (\ref{eq09}).
In particular, when the infinite matter is used to define
functions $\pi_0(\scalr)$, $\pi_1(\scalr)$, and $\pi_2(\scalr)$, like in the DME,
they parametrically depend on the Fermi momentum $k_F$.
By associating the local density $\rho(\bbox{R})$ with $k_F$, functions
 $\pi_0(\scalr)$, $\pi_1(\scalr)$, and $\pi_2(\scalr)$ become
dependent on $\rho(\bbox{R})$, and
hence the damping of the density matrix in the non-local direction
$\bbox{r}$ can be different in different local points $\bbox{R}$.
However, in order to keep the notation simple, we do not explicitly
indicate this possible dependence on density.

Within the quasi-local approximation, one obtains the exchange interaction energy,
   \be\label{eq19c}
   {\cal E}^{\text{int}}_{\text{exc}}
    = \int\rmd^3\bboxR \,{\cal H}^{\text{int}}_{\text{exc}}(\bboxR),
   \ee
where up to second order,
   \be\label{eq19d}
   {\cal H}^{\text{int}}_{\text{exc}}(\bboxR)
    = -\thalf \Big[ V^{00}_{\pi0} \rho^2 + \tfrac{1}{3} V^{02}_{\pi2}
                           \Big(\tfrac{1}{4}\rho\Delta\rho
              -(\rho\tau-\bbox{j}^{\,2})\Big)\Big] + \ldots,
   \ee
and where $\tau=\tau_{aa}$.
The coupling constants, $V^{00}_{\pi0}$ and $V^{02}_{\pi2}$, are given by the
following moments of the interaction,
   \be\label{eq11}
  V^{ij}_{\pi n} = \int\rmd^3\bboxr\,  r^n
                               \pi_i(\scalr) \pi_j(\scalr) V(r)
            =  4\pi\int\rmd r\,  r^{n+2}
                               \pi_i(\scalr) \pi_j(\scalr) V(r) .
   \ee
Unlike the coupling constants defining the direct term (\ref{eq02}),
these in Eq.~(\ref{eq11}) have to be understood as
the running coupling constants, which in the DME depend on the scale
of the Fermi momentum $k_F$ or density $\rho(\bbox{R})$.

Again, the separation of scales between the range of interaction and
the rate of change of the density matrix in the non-local direction
results in the dependence of the local energy density on two coupling
constants only, and not on the details of the interaction. For the
parabolic approximation of Eq.~(\ref{eq08}), the coupling constants
that define the direct and exchange energies are identical, i.e.,
$V^{00}_{\pi0}$=$V_0$ and $V^{02}_{\pi2}$=$V_2$; however, for the quasi-local approximation of
Eq.~(\ref{eq09}) they are different. This important observation is
discussed in Sec.~\ref{sec3b} in more detail.

In nuclei, the separation of scales discussed above is not very well
pronounced. The characteristic parameter, which defines these
relative scales, is equal to $k_Fa$, where $a$ stands for the range
of the interaction. For example, within the Gogny interaction, which we
analyze in detail in Sec.~\ref{sec4}, there are two components with
the ranges of $a=0.7$ and 1.2\,fm, whereas $k_F=1.35$\,fm$^{-1}$, which
gives values of $k_Fa=0.95$ and 1.62 that are dangerously close to 1.
Therefore, in the expansion of the local energy density
(\ref{eq19d}), one cannot really count on the moments $V^{ij}_{\pi n}$
decreasing with the increasing order $n$.

Instead, as demonstrated by Negele and Vautherin \cite{[Neg72]}, one
can hope for tremendously improving the convergence by using at each
order the proper counter-terms, which make each order vanish in the
infinite matter. Without repeating the original NV construction,
here we only note that the net result consists of adding and
subtracting in Eq.~(\ref{eq09}) the infinite-matter term, that is,
   \bn
   \rho(\bboxr_1,\bboxr_2)  &=&
                     \nu_0(\scalr)\rho(\bboxR)
                    +  i\nu_1(\scalr)  r_aj_a\rho(\bboxR)
\nonumber \\
              &+& \tfrac{1}{2} \nu_2(\scalr) r_a r_b
\left[\tfrac{1}{4}\nabla_a\nabla_b\rho(\bboxR)-\tau_{ab}(\bboxR)+
    \tfrac{1}{5}\delta_{ab}k_F^2\rho(\bboxR)\right] + \ldots~,
\label{eq09b}
   \en
where we have defined functions $\nu_i(r)$ such that:
   \be
   \nu_0(r)   =   \pi_0(r) - \tfrac{1}{10}(k_F r)^2\pi_2 , \quad\quad
   \nu_1(r)   =   \pi_1(r) , \quad\quad
   \nu_2(r)   =   \pi_2(r) .
\label{eq09c}
   \ee

By neglecting the term quadratic in $\nu_2$, we can now use the
approximation (\ref{eq09b}) to calculate the product of densities in
Eq.~(\ref{eq10}), which gives
   \bn\!\!\!\!\!\!\!\!\!\!\!\!\!\!\!\!\!\!\!\!\!\!\!\!\!\!\!\!\!\!\!\!
   \rho(\bboxr_1,\bboxr_2)\rho(\bboxr_2,\bboxr_1) &=&
                 \nu_0^2(\scalr)\rho^2(\bboxR)
               + \nu_0(\scalr)\nu_2(\scalr) r_a r_b
\Big\{\tfrac{1}{4}\rho(\bboxR)\nabla_a\nabla_b\rho(\bboxR)-\rho(\bboxR)\tau_{ab}(\bboxR)
\nonumber \\
              &&~~~~~~~~~~~~~~~~~~~~
   +\tfrac{1}{5}\delta_{ab}k_F^2\rho^2(\bboxR)
              +j_a(\bboxR)j_b(\bboxR)\Big\} + \ldots~,
\label{eq10a}
   \en
where again the gauge invariance requires that
\be
\nu_1^2(\scalr)=\nu_0(\scalr)\nu_2(\scalr).
\label{eq41}
\ee
We note that the gauge-invariance conditions for the functions $\pi_i(r)$
and $\nu_i(r)$, Eqs.~(\ref{eq40}) and (\ref{eq41}), are compatible
with one another only up to the term $\pi_2^2(\scalr)$, which was
shifted to higher orders.
Finally, approximation (\ref{eq10a}) gives the energy density
analogous to (\ref{eq19d}),
   \be\label{eq19e}
   {\cal H}^{\text{int}}_{\text{exc}}(\bboxR)
    = -\thalf \Big[ V^{00}_{\nu0} \rho^2 + \tfrac{1}{3} V^{02}_{\nu2}
                           \Big(\tfrac{1}{4}\rho\Delta\rho
              -(\rho\tau-\bbox{j}^{\,2})
   +\tfrac{3}{5}k_F^2\rho^2
\Big)\Big] + \ldots,
   \ee
where the coupling constants, $V^{00}_{\nu0}$ and $V^{02}_{\nu2}$, are given by the
moments of the interaction calculated for functions $\nu_i(r)$, namely,
   \be\label{eq11a}
  V^{ij}_{\nu n} = \int\rmd^3\bboxr\,  r^n
                               \nu_i(\scalr) \nu_j(\scalr) V(r)
            =  4\pi\int\rmd r\,  r^{n+2}
                               \nu_i(\scalr) \nu_j(\scalr) V(r) .
   \ee

We see that the two sets of auxiliary functions, $\pi_i(r)$ and
$\nu_i(r)$, are suitable for discussing the approximate forms of the
nonlocal density $\rho(\bboxr_1,\bboxr_2)$ and exchange energy
density ${\cal H}^{\text{int}}_{\text{exc}}(\bboxR)$, respectively.
Although for the nonlocal density they correspond to a simple
reshuffling of terms, which gives relations (\ref{eq09c}) between
$\pi_i(r)$ and $\nu_i(r)$, for the exchange energy density they
constitute entirely different approximations, given in
Eqs.~(\ref{eq19d}) and (\ref{eq19e}), with expansion (\ref{eq19e})
having a larger potential for faster convergence. Only this
latter expansion is further discussed.

\subsection{Determination of functions $\pi_i(r)$}\label{sec2c}

For a given nonlocal density $\rho(\bboxr_1,\bboxr_2)$, the auxiliary
functions $\pi_i(r)$ or $\nu_i(r)$,
which define its quasi-local approximation, can be calculated as their best possible
approximations in terms of local densities. However, the usefulness
of the expansion relies on the assumption that generic forms of these
functions can be estimated and then applied to all many-body systems
of a given kind.

The standard
Slater approximation \cite{[Sla51],[Bok99]}, which is routinely used to evaluate the
Coulomb exchange energy (cf.~Refs.~\cite{[Tit74],[Ska01]}), corresponds to
\be\label{eq13}
\pi_0(r) = \nu_0(r) = \frac{3j_1(k_Fr)}{k_Fr}  \quad\mbox{and}\quad
\pi_2(r) = 0 .
\ee
The NV expansion \cite{[Neg72]}
gives a second-order estimate by making the momentum expansion around the
Fermi momentum $k_F$ of an infinite system. This gives:
\bn
\pi_0(r) &=& \frac{6j_1(k_Fr)+21j_3(k_Fr)}{2k_Fr} \simeq 1 - \frac{(k_Fr)^4}{504},
\label{eq12a}
 \\
\pi_2(r) &=& \frac{105j_3(k_Fr)}{(k_Fr)^3}  \simeq 1 - \frac{(k_Fr)^2}{18}
                                                     + \frac{(k_Fr)^4}{792},
\label{eq12b}
 \\
\nu_0(r) &=& \frac{3j_1(k_Fr)}{k_Fr} \simeq 1 - \frac{(k_Fr)^2}{10}
                                                     + \frac{(k_Fr)^4}{280},
\label{eq12c}
\en
where $j_n(k_Fr)$ are the spherical Bessel functions.

\begin{figure}
\begin{center}
\includegraphics[width=0.6\columnwidth]{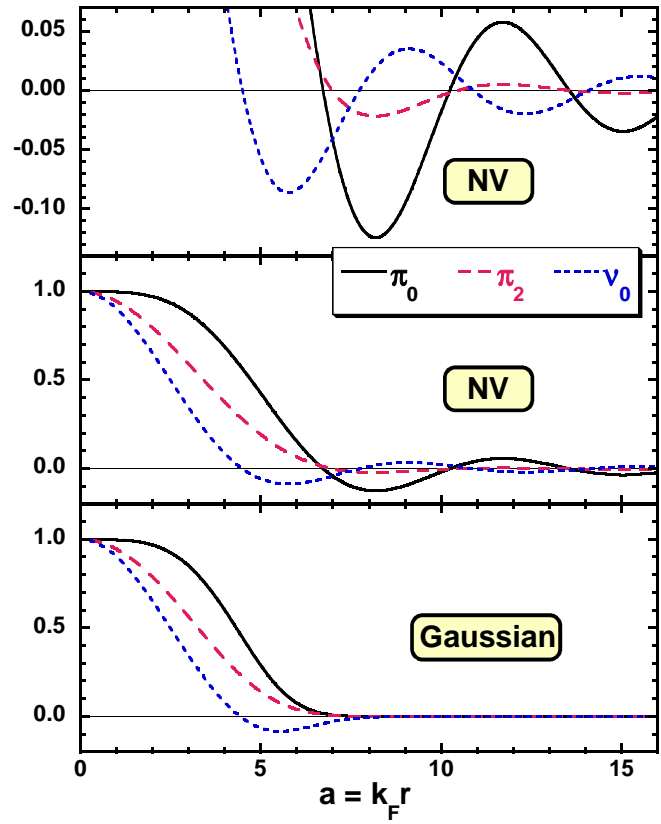}
\caption{\label{fig1}
Dependence of the functions $\pi_0(r)$ (solid lines), $\pi_2(r)$ (dashed lines),
and $\nu_0(r)$ (dotted lines) on $k_Fr$. The top and middle panels show
the NV functions of Eqs.~(\protect\ref{eq12a})--(\protect\ref{eq12c}),
with the top panel plotted in expanded scale to better show the details
at large $k_Fr$. The bottom panel shows the functions $\pi_0(r)$, $\pi_2(r)$,
and $\nu_0(r)$ approximated by Gaussians as in
Eqs.~(\protect\ref{eq121a})--(\protect\ref{eq121c}).
}
\end{center}
\end{figure}

The NV functions $\pi_0(r)$, $\pi_2(r)$, and $\nu_0(r)$ are plotted
in Fig.~\ref{fig1} with solid, dashed, and dotted lines,
respectively. One can see that for large $k_Fr$, functions $\pi_0(r)$
and $\pi_2(r)$ have zeros close to one another and the same signs.
Indeed, asymptotically both behave like $\cos(k_Fr)/k_Fr$. Therefore,
the gauge-invariance condition (\ref{eq40}) can be satisfied almost everywhere.
On the other hand, functions $\nu_0(r)$ and
$\nu_2(r)=\pi_2(r)$, have asymptotically opposite signs, and the
corresponding gauge-invariance condition (\ref{eq41}) can be almost
nowhere satisfied. Nevertheless, as discussed in Sec.~\ref{sec2b},
what really matters are the moments of interaction (\ref{eq11}) and
(\ref{eq11a}), where functions $\pi_i(r)$ and $\nu_i(r)$ are probed
only within the range of the interaction, that is, up to
$k_Fr\,\simeq\,$1--2. Therefore, for the NV expansion, one can safely
use approximations:
\bn
\pi_1(r) &=& +\sqrt{|\pi_0(r)\pi_2(r)|} ,
\label{eq12d}
 \\
\nu_1(r) &=& +\sqrt{|\nu_0(r)\nu_2(r)|} ,
\label{eq12e}
\en
which are valid up to the first zero of $j_3$ or $j_1$, respectively,
that is, up to $k_Fr\,\simeq\,$7.0 and 4.5.

By the same token, we can replace in Eqs.~(\ref{eq12a})--(\ref{eq12c})
the Bessel functions by Gaussians having the same leading-order
dependence on $k_Fr$, namely,
\bn
\pi_0(r) &=& \exp\left(- \frac{(k_Fr)^4}{504}\right),
\label{eq121a}
 \\
\pi_2(r) &=& \exp\left(-\frac{(k_Fr)^2}{18} - \frac{(k_Fr)^4}{1100}\right),
\label{eq121b}
 \\
\nu_0(r) &=& \exp\left(- \frac{(k_Fr)^4}{504}\right) - \frac{(k_Fr)^2}{10}
              \exp\left(-\frac{(k_Fr)^2}{18} - \frac{(k_Fr)^4}{1100}\right).
\label{eq121c}
\en
As seen in the bottom panel of Fig.~\ref{fig1}, in this way, in the
region of small $k_Fr$, one obtains a very good reproduction of the
NV functions $\pi_i(r)$ and $\nu_i(r)$.

\section{Local energy density for particles with spin and isospin}\label{sec3}

\subsection{Density matrix expansion with spin and isospin}\label{sec3a}

For nucleons, the density matrix
$\rho(\bboxr_1\sigma_1\tau_1,\bboxr_2\sigma_2\tau_2)$ depends not only on
positions $\bboxr_1$ and $\bboxr_2$ but also on spin $\sigma_1,\sigma_2=\pm1$
and isospin $\tau_1,\tau_2=\pm1$ coordinates. Since the strong two-body interaction
is assumed to be isospin and rotationally invariant, it is convenient to
represent the standard density matrix
$\rho(\bboxr_1\sigma_1\tau_1,\bboxr_2\sigma_2\tau_2)$
through nonlocal densities $\rho_{\mu k}(\bboxr_1,\bboxr_2)$ as:
\be\label{eq:302}
\rho(\bboxr_1\sigma_1\tau_1,\bboxr_2\sigma_2\tau_2)
= \tfrac{1}{4}\sum_{\mu=0,x,y,z} \sum_{k=0}^3 \rho_{\mu k}(\bboxr_1,\bboxr_2)
\langle\sigma_1|\sigma_\mu|\sigma_2\rangle
\langle  \tau_1|  \tau_k  |\tau_2  \rangle ,
\ee
where $\sigma_0$ ($\sigma_{x,y,z}$) and $\tau_0$ ($\tau_{1,2,3}$) are the unity
(Pauli) matrices in the spin and isospin coordinates, respectively.
For $\mu=0$ and $\mu=x,y,z$, the densities are scalars and vectors, respectively,
and for $k=0$ and $k=1,2,3$, they are isoscalars and isovectors, so altogether
the density matrix is split into the four standard spin-isospin channels.

For the direct term, we can proceed as in Sec.~\ref{sec2a}, by
making the Taylor expansions of local densities (at
$\bboxr_1=\bboxr_2$) in each spin-isospin channel; that is, similarly
as in Eqs.~(\ref{eq03a}) and (\ref{eq03b}), we have
   \bn
   \label{eq03c}
   \rho_{\mu k}(\bboxR \pm \thalf\bboxr) &=&
                    \rho_{\mu k}(\bboxR) \pm \thalf r_a\nabla_a\rho_{\mu k}(\bboxR)
                                + \tfrac{1}{8} r_a r_b
                                  \nabla_a\nabla_b\rho_{\mu k}(\bboxR) + \ldots \,.
   \en
For the exchange term, Sec.~\ref{sec2b},
the analogous Taylor expansions of nonlocal densities, similarly
as in Eq.~(\ref{eq08}), read
\bn
\label{eq08a}\hspace*{-2cm}
   \rho_{\mu k}(\bboxR,\pm\bboxr) &=&
                    \rho_{\mu k}(\bboxR)
                  \pm i r_aj_{\mu ak}(\bboxR)
                  + \tfrac{1}{2} r_a r_b
           \left[\tfrac{1}{4}\nabla_a\nabla_b\rho_{\mu k}(\bboxR)-\tau_{\mu abk}(\bboxR)\right] + \ldots \,,
\en
where the current ($j_{\mu ak}(\bboxR)$) and kinetic
($\tau_{\mu abk}(\bboxR)$) densities are defined in each channel as in
Eqs.~(\ref{eq38}), namely,
\be
\label{eq38a}
   j_{\mu ak}(\bboxR)   =\tfrac{1}{i}\partial_a\rho_{\mu k}(\bboxR,\bboxr)_{\bboxr=0}, \quad\quad
\tau_{\mu abk}(\bboxR)=\nabla_a^{(1)}\nabla_b^{(2)}\rho_{\mu k}(\bboxr_1,\bboxr_2)_{\bboxr_1=\bboxr_2} .
\ee

The local density approximation of densities in all channels, analogous to
Eqs.~(\ref{eq09}) and (\ref{eq09b}), is now postulated as
\bn
\label{eq09f}
\hspace*{-2cm}
   \rho_{\mu k}(\bboxR,\pm\bboxr) &=&
                  \pi_0(r)  \rho_{\mu k}(\bboxR)
                  \pm i\pi_1(r) r_aj_{\mu ak}(\bboxR)
\nonumber \\
                 &&+ \tfrac{1}{2}\pi_2(r) r_a r_b
           \left[\tfrac{1}{4}\nabla_a\nabla_b\rho_{\mu k}(\bboxR)-\tau_{\mu abk}(\bboxR)\right] + \ldots \,,
\en
and
\bn
\label{eq09e}
\hspace*{-2cm}
   \rho_{\mu k}(\bboxR,\pm\bboxr) &=&
                     \nu_0(\scalr)\rho_{\mu k}(\bboxR)
                  \pm i\nu_1(\scalr)  r_aj_{\mu ak}\rho(\bboxR)
\nonumber \\
              &&+ \tfrac{1}{2} \nu_2(\scalr) r_a r_b
\left[\tfrac{1}{4}\nabla_a\nabla_b\rho_{\mu k}(\bboxR)-\tau_{\mu abk}(\bboxR)+
    \tfrac{1}{5}\delta_{ab}k_F^2\rho_{\mu k}(\bboxR)\right] + \ldots~.
\en

At this point, we have assumed that functions $\pi_i(r)$ and
$\nu_i(r)$ are channel-independent, that is, that they are
scalar-isoscalar functions. In~\ref{appA} we discuss this point in
more detail, and we show that the postulate of simply
channel-dependent functions $\pi_i(r)$ and $\nu_i(r)$ is incompatible
with properties of infinite matter, whereas the proper treatment of
the problem leads immediately to the channel mixing and the energy density, which
is not invariant with respect to rotational and isospin symmetries. This
question certainly requires further study, whereas at the moment, a
consistent approach can only be obtained by assuming the
scalar-isoscalar functions $\pi_i(r)$ and $\nu_i(r)$.

We can now apply derivations presented in Secs.~\ref{sec2a} and \ref{sec2b}
to the general case of an arbitrary finite-range local nuclear interaction
composed of the standard central, spin-orbit, and tensor terms:
   \bn
    \hat V(\bboxr_1,\bboxr_2)&=&W(r)
                              + B(r) P_\sigma
                              - H(r) P_\tau
                              - M(r) P_\sigma P_\tau
\nonumber \\
                        &+&\Big[P(r)
                         +      Q(r) P_\tau\Big]\bbox{L}\cdot\bbox{S}
                         + \Big[R(r)
                         +      S(r) P_\tau\Big]S_{12},
    \en
where $r$=$|\bboxr|$=$|\bboxr_1$$-$$\bboxr_2|$, and
\be
P_\sigma=\thalf(1+\bbox{\sigma}_1\cdot\bbox{\sigma}_2), \quad\quad
P_\tau  =\thalf(1+\vec {\tau  }_1\circ\vec {\tau  }_2),
\ee
\be
\bbox{L}=-i\hbar\bboxr\times\bbox{\partial}, \quad\quad
\bbox{S}=\tfrac{\hbar}{2}(\bbox{\sigma}_1+\bbox{\sigma}_2),  \quad\quad
  S_{12}=\tfrac{3}{r^2}(\bbox{\sigma}_1\cdot\bboxr)(\bbox{\sigma}_2\cdot\bboxr)-
\bbox{\sigma}_1\cdot\bbox{\sigma}_2.
\ee
After
straightforward but lengthy calculations, one obtains the interaction energy in
the form of a local integral, analogous to that for the Skyrme
interaction \cite{[Flo75],[Eng75],[Per04]},
\bn
   &&{\cal E}^{\text{int}} = \int\rmd^3\bboxR \sum_{k=0}^3\Bigg[
      C_t^{\rho}            \rho_k^2
    + C_t^{\Delta\rho}      \rho_k\Delta\rho_k
    + C_t^{\tau}       \Big(\rho_k\tau_k
    -                       \bbox{j}_k^{\,2}\Big)
\nonumber \\
&& ~~~~~~~~~~~~~~~~~~~~~
    + C_t^{   s}            \bbox{s}_k^{\,2}
    + C_t^{\Delta s}        \bbox{s}_k\cdot\Delta
                            \bbox{s}_k
    + C_t^{   T}       \Big(\bbox{s}_k\cdot\bbox{T}_k
                          - {\mathsf J}_{abk} {\mathsf J}_{abk}\Big)
\nonumber \\
&& ~~~~~~~~~~~~~~~~~~~~~
    + C_t^{F}          \Big(\bbox{s}_k   \cdot\bbox{F}_k
                    -  \thalf {\mathsf J}_{aak} {\mathsf J}_{bbk}
                    -  \thalf {\mathsf J}_{abk} {\mathsf J}_{bak}\Big)
    + C_t^{\nabla{s}}  \Big(\bbox{\nabla}\cdot\bbox{s}_k\Big)^2
\nonumber \\
&& ~~~~~~~~~~~~~~~~~~~~~
    + C_t^{\nabla{J}}       \Big(\rho_k\bbox{\nabla}\cdot\bbox{J}_k
    +                       \bbox{s}_k\cdot(\bbox{\nabla}\times\bbox{j}_k)\Big)      \Bigg],
\label{eq18}
   \en
\ew
where
$\rho_k                 \equiv\rho_{0k}$,
$\tau_k                 \equiv\tau_{0bbk}$,
$\bbox{j}_{ak}          \equiv j_{0ak}$,
$\bbox{s}_{ak}          \equiv \rho_{ak}$,
$\bbox{T}_{ak}          \equiv\tau_{abbk}$,
$\bbox{F}_{ak}          \equiv\thalf(\tau_{babk}+\tau_{bbak})$,
${\mathsf J}_{abk}      \equiv j_{abk}$, and
$\bbox{J}_{ak}          \equiv \epsilon_{abc}j_{cbk}$
are the standard local densities. The isoscalar ($t=0$) and isovector
($t=1$) coupling constants $C_t$ correspond
to $k=0$ and $k=1,2,3$, respectively.

The coupling constants of the local
energy density (\ref{eq18}) are related to moments of the interaction
in the following way:
\newlength{\sep}
\setlength{\sep}{0.2em}
   \bn
\label{eq20}\hspace*{-2cm}
   8\left(\ba{l}C_0^{\rho}\\
                C_1^{\rho}\\
                C_0^{   s}\\
                C_1^{   s}\ea\right)
 &=&\left(\ba{r@{\hspace{\sep}}r@{\hspace{\sep}}r@{\hspace{\sep}}r}
                    4  &\pr2 & -2 & -1\\
                    0  &   0 & -2 & -1\\
                    0  &   2 &  0 & -1\\
                    0  &   0 &  0 & -1\ea\right)
    \left(\ba{c}W_0+M^{00}_{\nu0}+\tfrac{1}{5}M^{02}_{\nu2}\,k_F^2\\
                B_0+H^{00}_{\nu0}+\tfrac{1}{5}H^{02}_{\nu2}\,k_F^2\\
                H_0+B^{00}_{\nu0}+\tfrac{1}{5}B^{02}_{\nu2}\,k_F^2\\
                M_0+W^{00}_{\nu0}+\tfrac{1}{5}W^{02}_{\nu2}\,k_F^2
   \ea\right) ,
\\[2ex]
\label{eq21}\hspace*{-2cm}
  96\left(\ba{l}C_0^{\Delta\rho}\\
                C_1^{\Delta\rho}\\
                C_0^{\tau}      \\
                C_1^{\tau}      \\
                C_0^{\Delta s}  \\
                C_1^{\Delta s}  \\
                C_0^{   T}      \\
                C_1^{   T}      \\
                C_0^{   F}      \\
                C_1^{   F}      \\
                C_0^{\nabla s}  \\
                C_1^{\nabla s}
\ea\right)
 &=&
\left(\ba{r@{\hspace{\sep}}r@{\hspace{\sep}}r@{\hspace{\sep}}r@{\hspace{\sep}}r@{\hspace{\sep}}r@{\hspace{\sep}}r@{\hspace{\sep}}r@{\hspace{\sep}}r@{\hspace{\sep}}r@{\hspace{\sep}}r@{\hspace{\sep}}r}
                    8 & -1 &\pr4 & -2 & -4 &  2 & -2 &  4 &\pr0 &\pr0 &\pr0 &\pr0 \\
                    0 & -1 &   0 & -2 & -4 &  0 & -2 &  0 &   0 &   0 &   0 &   0 \\
                    0 &  4 &   0 &  8 &  0 & -8 &  0 &-16 &   0 &   0 &   0 &   0 \\
                    0 &  4 &   0 &  8 &  0 &  0 &  0 &  0 &   0 &   0 &   0 &   0 \\
                    0 & -1 &   4 &  0 &  0 &  2 & -2 &  0 &  -4 &   1 &  -2 &   2 \\
                    0 & -1 &   0 &  0 &  0 &  0 & -2 &  0 &   0 &   1 &  -2 &   0 \\
                    0 &  4 &   0 &  0 &  0 & -8 &  0 &  0 &   0 &  -4 &   0 &  -8 \\
                    0 &  4 &   0 &  0 &  0 &  0 &  0 &  0 &   0 &  -4 &   0 &   0 \\
                    0 &  0 &   0 &  0 &  0 &  0 &  0 &  0 &   0 &  12 &   0 &  24 \\
                    0 &  0 &   0 &  0 &  0 &  0 &  0 &  0 &   0 &  12 &   0 &   0 \\
                    0 &  0 &   0 &  0 &  0 &  0 &  0 &  0 & -12 &   3 &  -6 &   6 \\
                    0 &  0 &   0 &  0 &  0 &  0 &  0 &  0 &   0 &   3 &  -6 &   0
\ea\right)
    \left(\ba{l} {\phac}W     _2 \\
                 {\phac}W^{02}_{\nu2} \\
                 {\phac}B     _2      \\
                 {\phac}B^{02}_{\nu2} \\
                 {\phac}H     _2      \\
                 {\phac}H^{02}_{\nu2} \\
                 {\phac}M     _2      \\
                 {\phac}M^{02}_{\nu2} \\
                 {\prac}R     _2      \\
                 {\prac}R^{02}_{\nu2} \\
                 {\prac}S     _2      \\
                 {\prac}S^{02}_{\nu2}
\ea\right) ,
\\[2ex]
\label{eq22}\hspace*{-2cm}
   24\left(\ba{l}C_0^{\nabla{J}}\\
                C_1^{\nabla{J}}\ea\right)
 &=&\left(\ba{rr}   2  & 1 \\
                    0  & 1 \ea\right)
    \left(\ba{c}P_2+Q^{01}_{\nu2}\\
                Q_2+P^{01}_{\nu2}
   \ea\right) .
   \en
All the coupling constants of the local energy density (\ref{eq18})
depend linearly on the following moments of potentials:
   \bnll{eq15}
\label{eq29a}
   X_n      &=& \int\rmd^3\bboxr\,  r^{\,n} X(r)
             =  4\pi\int\rmd\,r\,   r^{\,n+2} X(r) ,  \\
\label{eq29b}
   X^{ij}_{\nu n} &=& \int\rmd^3\bboxr\,  r^{\,n}  \nu_i(\scalr)\nu_j(\scalr) X(r) ,
                   =  4\pi\int\rmd\,r\,   r^{\,n+2}\nu_i(\scalr)\nu_j(\scalr) X(r) ,
   \enll
where $X$ stands for $W$, $B$, $H$, $M$, $P$, $Q$, $R$, or $S$.

Again we see that whenever expansions of density matrices,
Eqs.~(\ref{eq03c}) and (\ref{eq09e}), are sufficiently accurate within
the ranges of interactions, information about these interactions
collapses to a few lowest moments. Short-range details of these
interactions are, therefore, entirely irrelevant for low-energy
characteristics of nuclear states. This is typical of all
physical situations, where scales of interaction and observation are
well separated, as specified in the effective field theories. The
energy density characterizing the low-energy effects is local and
depends on local densities and their derivatives up to second order,
whereas the dynamic information is contained in a few coupling
constants.

Moreover, the detailed large-$r$ dependence of auxiliary functions
$\nu_i(\scalr)$ on position $r$ is also irrelevant, because all that
matters are moments (\ref{eq29b}) which define the coupling
constants (\ref{eq20})--(\ref{eq22}) describing the exchange energy, and these are influenced
only by the small-$r$ properties of functions
$\nu_i(\scalr)$. Finally, the most important feature is the $k_F$ or
density dependence of $\nu_i(\scalr)$, which determines the density
dependence of the coupling constants.

\subsection{Local energy density corresponding to the Skyrme
force}\label{sec3b}

In general, the number of moments entering
Eqs.~(\ref{eq20})--(\ref{eq22}) is higher than the number of
coupling constants, and all the coupling constants are independent.
However, it is extremely instructive to check what happens
in the vacuum limit of $k_F=0$. This situation is obtained by
setting $\nu_i(\scalr)=1$, which gives the direct and exchange
moments equal to one another, namely, $X_{\nu n}^{ij}$=$X_n$,
and the coupling constants of Eqs.~(\ref{eq20})--(\ref{eq22}) collapse to:
   \bn
\label{eq20a}
   8\left(\ba{l}C_0^{\rho}\\
                C_1^{\rho}\\
                C_0^{   s}\\
                C_1^{   s}\ea\right)
 &=&\left(\ba{r@{\hspace{\sep}}r}
                    3  &\pr0  \\
                   -1  &  -2  \\
                   -1  &   2  \\
                   -1  &   0  \ea\right)
    \left(\ba{c}W_0+M_0\\
                B_0+H_0
   \ea\right) ,
\\[2ex]
\label{eq21a}
  96\left(\ba{l}C_0^{\Delta\rho}\\
                C_1^{\Delta\rho}\\
                C_0^{\tau}      \\
                C_1^{\tau}      \\
                C_0^{\Delta s}  \\
                C_1^{\Delta s}  \\
                C_0^{   T}      \\
                C_1^{   T}      \\
                C_0^{   F}      \\
                C_1^{   F}      \\
                C_0^{\nabla s}  \\
                C_1^{\nabla s}
\ea\right)
 &=&
    \left(\ba{r@{\hspace{\sep}}r@{\hspace{\sep}}r@{\hspace{\sep}}r@{\hspace{\sep}}r@{\hspace{\sep}}r}
                         7 &\pr2 & -2 &  2 &\pr0 &\pr0 \\
                        -1 &  -2 & -4 & -2 &   0 &   0 \\
                         4 &   8 & -8 &-16 &   0 &   0 \\
                         4 &   8 &  0 &  0 &   0 &   0 \\
                        -1 &   4 &  2 & -2 &  -3 &   0 \\
                        -1 &   0 &  0 & -2 &   1 &  -2 \\
                         4 &   0 & -8 &  0 &  -4 &  -8 \\
                         4 &   0 &  0 &  0 &  -4 &   0 \\
                         0 &   0 &  0 &  0 &  12 &  24 \\
                         0 &   0 &  0 &  0 &  12 &   0 \\
                         0 &   0 &  0 &  0 &  -9 &   0 \\
                         0 &   0 &  0 &  0 &   3 &  -6
\ea\right)
    \left(\ba{l} {\phac}W     _2 \\
                 {\phac}B     _2\\
                 {\phac}H     _2\\
                 {\phac}M     _2\\
                 {\prac}R     _2\\
                 {\prac}S     _2
\ea\right) ,
\\[2ex]
\label{eq22a}
   24\left(\ba{l}C_0^{\nabla{J}}\\
                C_1^{\nabla{J}}\ea\right)
 &=&\left(\ba{r}    3 \\
                    1 \ea\right)
    \left(\ba{c}P_2+Q_2
   \ea\right) .
   \en
These coupling constants correspond exactly to those
obtained for the Skyrme force (see Ref.~\cite{[Per04]} for the
notations and conventions used), namely,
\bn
\label{eq4.76a}
t_0    ~= \phantom{-}              W_0 + M_0  \quad&,&\quad
t_0x_0  = \phantom{-}              B_0 + H_0 , \\
\label{eq4.76b}
t_1    ~=         {-}\tfrac{1}{3} (W_2 + M_2) \quad&,&\quad
t_1x_1  =         {-}\tfrac{1}{3} (B_2 + H_2), \\
\label{eq4.76c}
t_2    ~= \phantom{-}\tfrac{1}{3} (W_2 - M_2) \quad&,&\quad
t_2x_2  = \phantom{-}\tfrac{1}{3} (B_2 - H_2), \\
\label{eq4.76d}
t_e    ~= \phantom{-}\tfrac{1}{15}(S_2 - R_2) \quad&,&\quad
t_o  ~~~= \phantom{-}\tfrac{1}{15}(S_2 + R_2), \\
\label{eq4.76e}
W       =         {-}\tfrac{1}{6} (P_2 + Q_2) \quad&.&
\en

The same relations are also obtained by using in the exchange term
the pure Taylor expansions (\ref{eq08a}); that is, by setting
$\pi(r)=1$, which gives $X_{\pi n}^{ij}$=$X_n$, and by using the
classification of terms as in Eqs.~(\ref{eq19c})--(\ref{eq11}). This
second way of obtaining the approximate coupling constants leads to
results independent of $k_F$, which are, of course, identical to
those obtained at $k_F=0$ above.

Relations (\ref{eq20a})--(\ref{eq22a}) imply that the coupling
constants of the energy functional (\ref{eq18}) are dependent of one
another, and in fact, half of them determines the other half. This
is exactly the situation encountered when the energy density is
calculated for the Skyrme interaction. Then one obtains
(cf.\ Ref.~\cite{[Dob95e]}):
\bn\hspace*{0.3cm}
\label{eq4.74a}
   3\left(\ba{l}C_0^{   s}\\
                C_1^{   s}\ea\right)
   &=&\left(\ba{rr}   -2 & -3 \\
                    -1 &  0 \ea\right)
                \left(\ba{c}C_0^{\rho}\\
                            C_1^{\rho}\ea\right) ,
\\[2ex]
\label{eq4.69a}
  24\left(\ba{l}C_0^{\Delta s}  \\
                C_1^{\Delta s}  \\
                C_0^{   T}      \\
                C_1^{   T}      \\
                C_0^{\nabla s}  \\
                C_1^{\nabla s}
\ea\right)
   &=&\left(\ba{rrrrrr}
                       -12  &   -12  &     3  &     9   &   0 &  -6 \\
                        -4  &    -4  &     3  &    -3   &  -2 &   4 \\
                        16  &    48  &    -4  &    12   &  -8 &   0 \\
                        16  &   -16  &     4  &   -12   &   0 &  -8 \\
                         0  &     0  &     0  &     0   &   0 & -18 \\
                         0  &     0  &     0  &     0   &  -6 &  12
\ea\right)
    \left(\ba{l}C_0^{\Delta\rho}\\
                C_1^{\Delta\rho}\\
                C_0^{\tau}      \\
                C_1^{\tau}      \\
                C_0^{   F}      \\
                C_1^{   F}
\ea\right) ,
\\[2ex]
\label{eq28}\hspace*{1.5cm}
                C_0^{\nabla{J}}
            &=&3C_1^{\nabla{J}} .
\en
It is obvious that the above relations among the coupling
constants result from an oversimplified approximation to the exchange
energy of the finite-range interaction.

We recall here \cite{[Dob95e],[Dob96b]} that without the tensor
terms, relations (\ref{eq4.74a}) and (\ref{eq4.69a}) allow us to
determine the time-odd coupling constants $C_t^{s}$, $C_t^{\Delta
s}$, and $C_t^{   T}$ as functions of the time-even coupling
constants $C_t^{\rho}$, $C_t^{\Delta\rho}$, and $C_t^{\tau}$. Since
the time-even coupling constants are usually adjusted solely to the
time-even observables, the resulting values of the time-odd coupling
constants are simply ``fictitious'' or ``illusory'', as noted already
in Ref.~\cite{[Neg75]}. In a more realistic case of relations
(\ref{eq20}) and (\ref{eq21}), these constraints are no longer valid,
and the time-odd properties of the functional are independent
of the time-even properties. This independence requires
breaking the link between the Skyrme force and the density
functional.

\section{Application to the Gogny interaction}\label{sec4}

In this section, we apply the results of Sec.~\ref{sec3} to the
finite-range part of the Gogny interaction D1S~\cite{[Ber91b]}. This
amounts to calculating moments (\ref{eq29a}) and (\ref{eq29b}) of
the Gaussian functions with the two ranges of 0.7 and 1.2\,fm,
which constitute the central part of the Gogny interaction. Because this
interaction does not contain any finite-range spin-orbit or tensor
force in Eqs.~(\ref{eq21}) and (\ref{eq22}), the moments $P$, $Q$,
$R$, and $S$ are set to zero. On the other hand, the zero-range
spin-orbit and density-dependent terms of the Gogny interaction are
left unchanged.

\begin{table}
\begin{center}
\caption{\label{tab1}
The NV coupling constants
(\protect\ref{eq20})--(\protect\ref{eq21}) calculated for the Gogny
interaction D1S~\protect\cite{[Ber91b]} for the Fermi momenta of  $k_F=0$ and 1.35\,fm$^{-1}$.
First-order coupling constants, $C_t^{\rho}$
and $C_t^{\Delta s}$, are in units of MeV\,fm$^3$ and the
second-order coupling constants, $C_t^{\Delta\rho}$, $C_t^{\tau}$,
$C_t^{\Delta s}$, and $C_t^{T}$, are in units of MeV\,fm$^5$.
}
\vspace*{1ex}
\begin{tabular}{|l|rr|rr|}
\cline{2-5} \multicolumn{1}{l|}{}
                   &  \multicolumn{2}{c|}{$t=0$}  &  \multicolumn{2}{c|}{$t=1$}     \\
\cline{2-5} \multicolumn{1}{l|}{}
                   &  \multicolumn{1}{c}{$k_F=0$} &  \multicolumn{1}{c|}{$k_F=1.35$}
                   &  \multicolumn{1}{c}{$k_F=0$} &  \multicolumn{1}{c|}{$k_F=1.35$} \\
\hline
$C_t^{\rho}      $ & $-$665.1658   &  $-$600.6156
                   & $ $468.5360   &  $ $428.3580  \\
$C_t^{   s}      $ & $-$25.09219   &  $-$57.13246
                   & $ $221.7219   &  $ $230.3318  \\
$C_t^{\Delta\rho}$ & $-$125.3365   &  $-$84.66327
                   & $ $56.65570   &  $ $31.22042  \\
$C_t^{\tau}      $ & $ $236.9227   &  $ $74.22964
                   & $-$141.9368   &  $-$40.19573  \\
$C_t^{\Delta s}  $ & $ $10.72944   &  $-$10.25276
                   & $ $58.80425   &  $ $65.14281  \\
$C_t^{   T}      $ & $-$80.70182   &  $ $3.226982
                   & $-$10.87263   &  $-$36.22687  \\
\hline
\end{tabular}
\end{center}
\end{table}

\begin{table}
\begin{center}
\caption{\label{tab2}
The standard Skyrme-force parameters
(\protect\ref{eq4.76a})--(\protect\ref{eq4.76c}) calculated for the
Gogny interaction D1S~\protect\cite{[Ber91b]} for the Fermi momenta of  $k_F=0$ and
1.35\,fm$^{-1}$. These parameters correspond to the time-even sector
of the Skyrme functional.
Parameter $t_0$ is in units of MeV\,fm$^3$;
parameters $t_1$ and $t_2$ are in units of MeV\,fm$^5$, and
parameters $x_0$, $x_1$, and $x_2$ are dimensionless.
}
\vspace*{1ex}
\begin{tabular}{|l|rr|}
\cline{2-3} \multicolumn{1}{l|}{}
         &  \multicolumn{1}{c}{$k_F=0$} &  \multicolumn{1}{c|}{$k_F=1.35$} \\
\hline
$t_0   $ & $-$1773.775    &  $-$1601.642  \\
$t_1   $ & $ $984.3584    &  $ $550.5103  \\
$t_2   $ & $ $810.3964    &  $-$166.0710  \\
$x_0   $ & $ $0.5565848   &  $ $0.5697973 \\
$x_1   $ & $ $0.2488322   &  $ $0.09972511\\
$x_2   $ & $-$0.9915809   &  $-$0.5517191 \\
\hline
\end{tabular}
\end{center}
\end{table}

\begin{figure}
\begin{center}
\includegraphics[width=0.8\columnwidth]{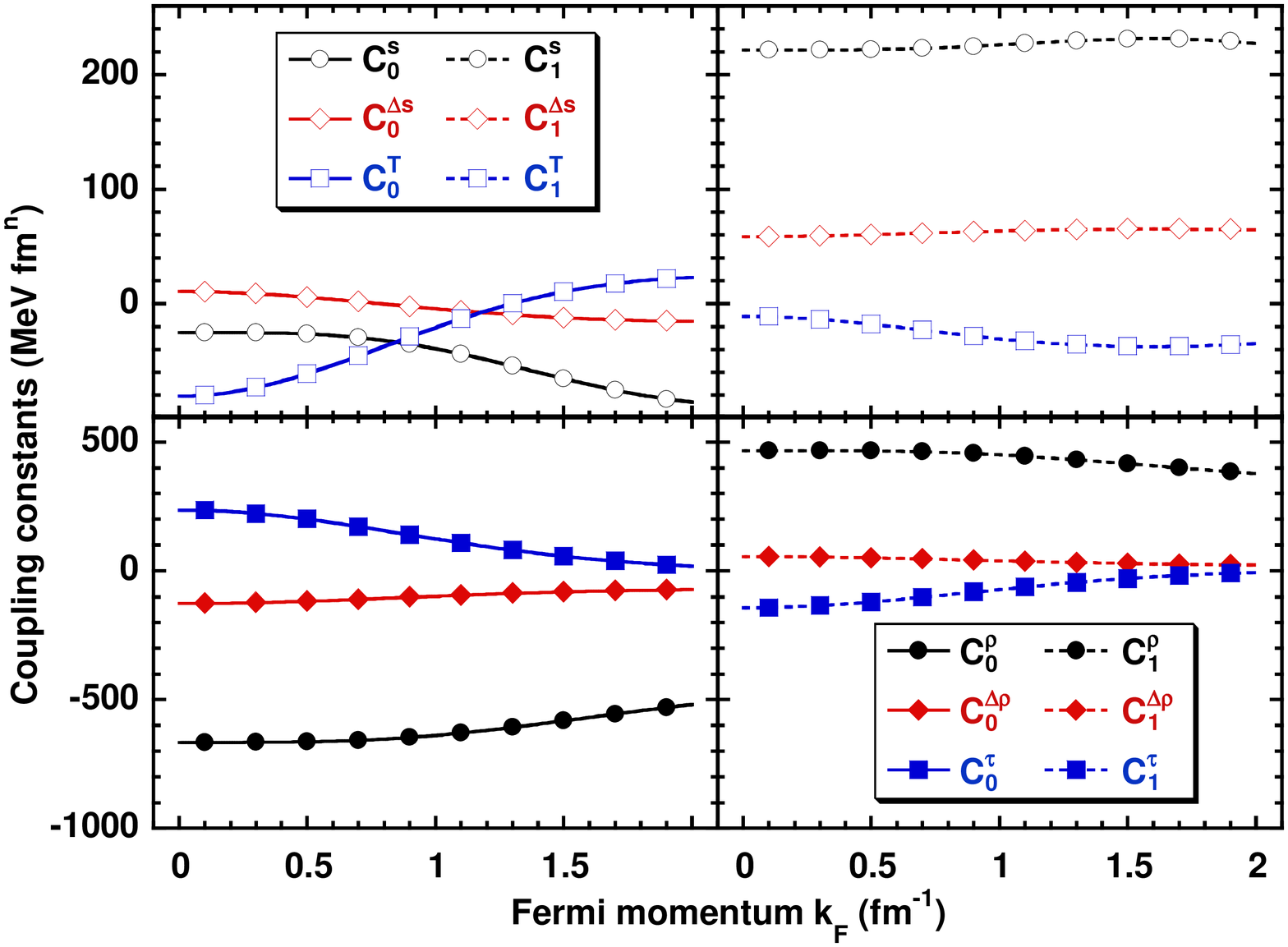}
\caption{\label{fig4}
Dependence of the NV coupling constants
(\protect\ref{eq20})--(\protect\ref{eq21}), calculated for the Gogny
interaction D1S~\protect\cite{[Ber91b]}, on the Fermi momentum $k_F$. Full and open symbols
(lower and upper panels) show values of the time-even and time-odd
coupling constants, respectively. Solid and dashed lines (left and
right panels) show values of the isoscalar and isovector coupling
constants, respectively. Units are specified in the caption to
Table~\protect\ref{tab1}.
}
\end{center}
\end{figure}

\begin{figure}
\begin{center}
\includegraphics[width=0.8\columnwidth]{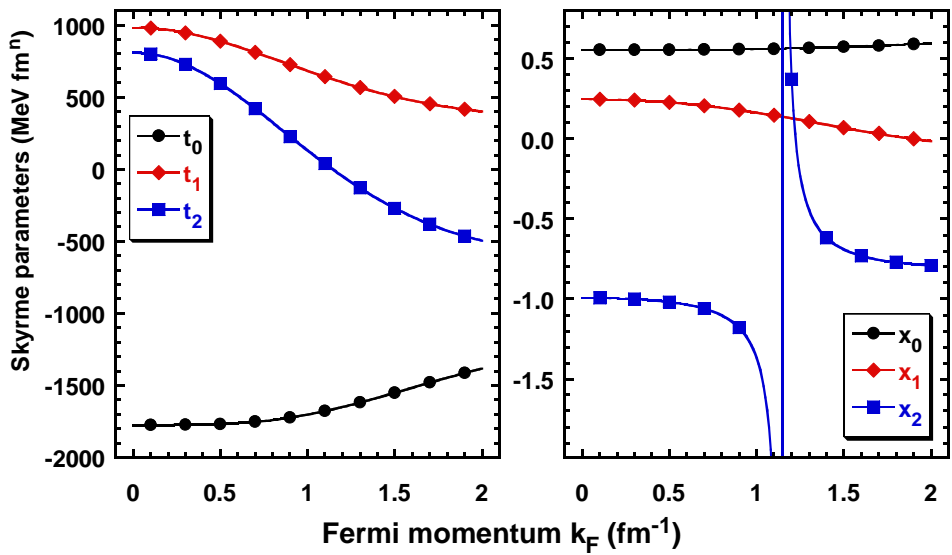}
\caption{\label{fig6}
Dependence of the standard Skyrme-force parameters
(\protect\ref{eq4.76a})--(\protect\ref{eq4.76c}), calculated for the Gogny
interaction D1S~\protect\cite{[Ber91b]}, on the Fermi momentum $k_F$. These parameters correspond
to the time-even sector of the Skyrme functional.
Units are specified in the caption of
Table~\protect\ref{tab2}.
}
\end{center}
\end{figure}

In Table~\ref{tab1} we show values of coupling constants
(\ref{eq20})--(\ref{eq22}) calculated in the vacuum ($k_F=0$) and at
the saturation density ($k_F=1.35$\,fm$^{-1}$). Similarly,
Table~\ref{tab2} shows values of the Skyrme-force parameters
(\ref{eq4.76a})--(\ref{eq4.76c}) corresponding to the time-even
coupling constants. In Figs.~\ref{fig4} and \ref{fig6} we plot
the coupling constants and Skyrme-force
parameters, respectively, as functions of the Fermi momentum.

The most important observation resulting from values shown in
Tables~\ref{tab1} and \ref{tab2} and Figs.~\ref{fig4} and \ref{fig6}
pertains to a significant density (or $k_F$) dependence of the
coupling constants and Skyrme-force parameters. The strongest
dependence is obtained for the isoscalar tensor coupling constant
$C_0^{T}$. (Note that the central finite-range Gogny interaction
induces significant values of the tensor coupling constants
$C_t^{T}$, even if this force does not contain any explicit tensor
term.) Also the kinetic coupling constants $C_t^{\tau}$ exhibit a
strong density dependence, going almost to zero at $k_F\sim2$\,fm$^{-1}$.
We note that the obtained density dependencies do not, in general, follow
any power laws. Significantly stronger density dependencies are
obtained for the Skyrme-force parameters (Fig.~\ref{fig6}). The pole
appearing in the parameter $x_2$ is a consequence of the fact
that parameters $t_2$ and $t_2x_2$, derived in Eq.~(\ref{eq4.76c}),
change signs at slightly different values of the Fermi momentum.

\begin{figure}
\begin{center}
\includegraphics[width=0.8\columnwidth]{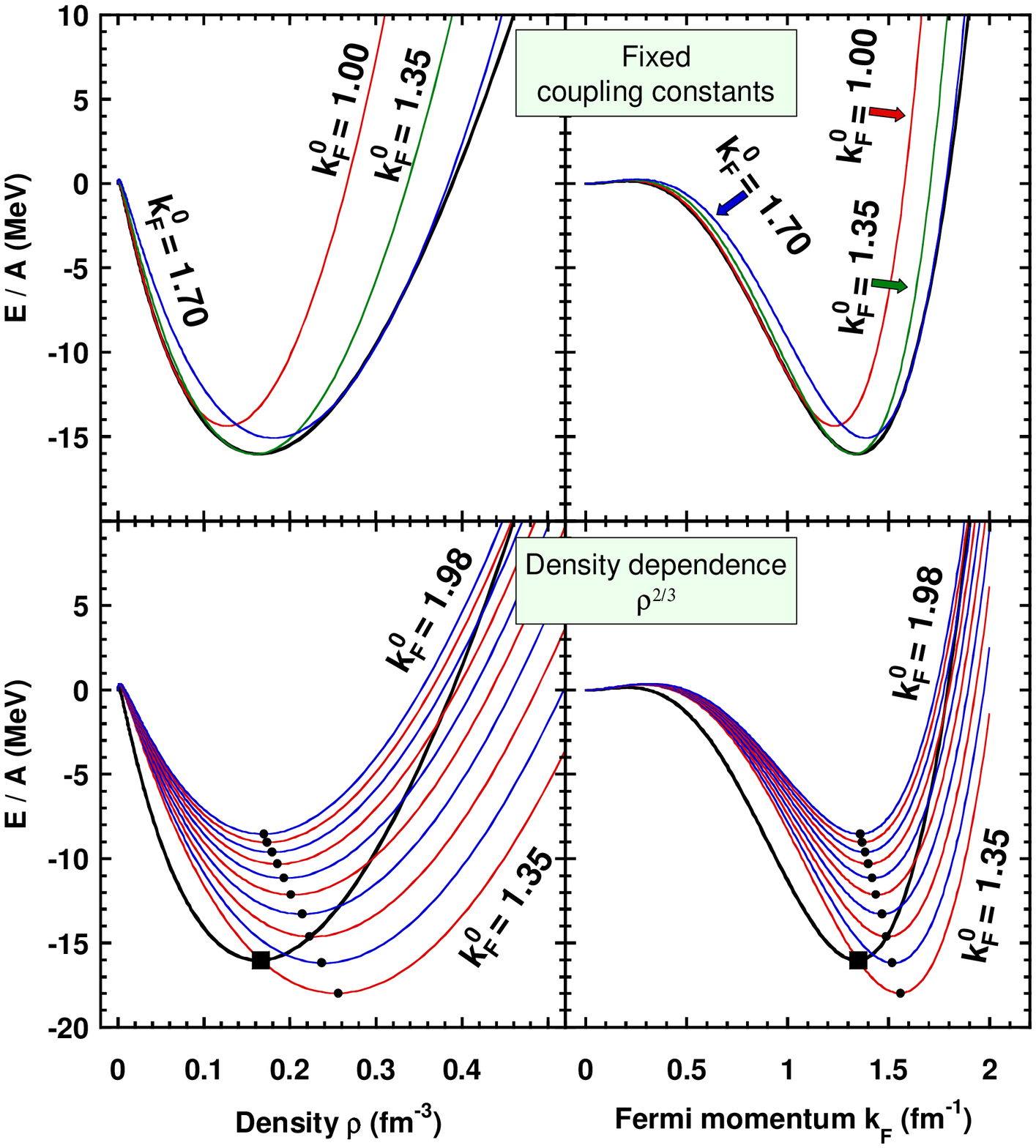}
\caption{\label{fig3}
Dependence of the infinite-matter energy per particle on the density
(left panels) or Fermi momentum (right panels). Thick lines show the NV
results for density-dependent coupling constants.
Thin lines show the NV results for fixed coupling constants
calculated at the given values of $k_F^0$. Upper and lower panels
show results for fixed coupling constants (\protect\ref{eq:313a})
and for coupling constants (\protect\ref{eq:313b}) that
depend on density as $\rho^{2/3}$. The full square and circles mark
the minima of curves.
}
\end{center}
\end{figure}

Let us now consider the question of whether parameters calculated at any
fixed value of the Fermi momentum can provide a reasonable
alternative. To analyze this point, we first note that the zero-order
coupling constants given in Eq.~(\ref{eq20}) depend on $k_F$ (i)
implicitly through the moments of Eq.~(\ref{eq29b}) and (ii)
explicitly through the $k_F^2$ term, that is,
\bn
C_t^{\rho}(k_F) &=& C_{t,0}^{\rho}(k_F) + C_{t,2/3}^{\rho}(k_F)\rho^{2/3} , \\
C_t^{s   }(k_F) &=& C_{t,0}^{s   }(k_F) + C_{t,2/3}^{s   }(k_F)\rho^{2/3} ,
\en
where we employed the standard association of the Fermi energy with density,
namely, $\rho=2k_F^3/3\pi^2$.
Therefore, we can consider fixed values of the coupling constants
calculated at a given Fermi momentum $k_F^0$ as
\bn
\label{eq:313a}
&&C_t^{\rho}(k_F^0) \mbox{~and~} C_t^{s   }(k_F^0),  \mbox{~or~}\\
\label{eq:313b}
&&C_{t,0}^{\rho}(k_F^0), C_{t,2/3}^{\rho}(k_F^0),
C_{t,0}^{s   }(k_F^0), \mbox{~and~} C_{t,2/3}^{s   }(k_F^0).
\en
The second option gives the coupling constants that still depend on
the density as $\rho^{2/3}$, in analogy with the standard
density-dependent term, which for the Gogny force depends on the
density as $\rho^{1/3}$, and which in the present study is alway kept
untouched.

In the upper and lower panels of Fig.~\ref{fig3}, thin lines show the
nuclear matter equations of state (energy per particle in function of
density or Fermi momentum) obtained for the coupling constants fixed
according to prescriptions (\ref{eq:313a}) and (\ref{eq:313b}),
respectively. For comparison, thick lines show the NV results
obtained for density-dependent coupling constants. Since in the
nuclear matter, the factor multiplying the term
$\nu_0(\scalr)\nu_2(\scalr)$ in Eq.~(\ref{eq10}) vanishes exactly (by
construction), the thick lines correspond to the exact Gogny force
results.

Equations of state calculated with prescription (\ref{eq:313b}) for
fixed values of $k_F^0$ from 1.35 to 1.98\,fm$^{-1}$ with the step of
0.07\,fm$^{-1}$ completely miss the saturation point. This means that
Skyrme forces with constant parameters (\ref{eq:313b}) derived by the
NV expansion cannot be equivalent to the finite-range Gogny
interaction. On the other hand, prescription (\ref{eq:313a}), for
fixed value of $k_F^0=1.35$\,fm$^{-1}$, reproduces the equation of
state fairly well, with some deviations seen only at densities beyond
the saturation point. For comparison, we also show results obtained
with $k_F^0=1$ and 1.7\,fm$^{-1}$, which fit the equation of state at
low and high densities, respectively. These results show that the
explicit dependence on the Fermi momentum, which appears in the NV
expansion, cannot be used to define the density dependence of the
coupling constants.

\begin{figure}
\begin{center}
\includegraphics[width=0.8\columnwidth]{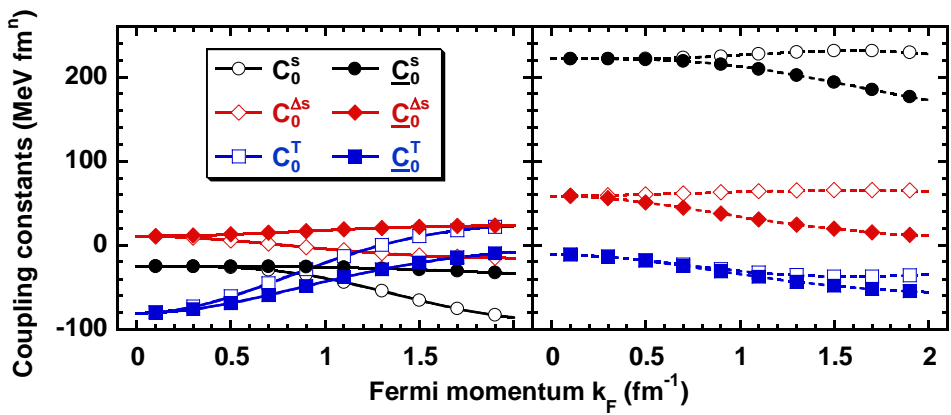}
\caption{\label{fig5}
Similar to Fig.~\protect\ref{fig4} except for the time-odd NV coupling
constants only. The open symbols show exact results
(\protect\ref{eq20})--(\protect\ref{eq21}) as plotted in
Fig.~\protect\ref{fig4}, whereas the full symbols show results inferred
from the time-even sector by using Eqs.~(\protect\ref{eq4.74a}) and
(\protect\ref{eq4.69a}). Solid and dashed lines (left and
right panels) show values of the isoscalar and isovector coupling
constants, respectively.
}
\end{center}
\end{figure}

\begin{table}
\begin{center}
\caption{\label{tab3}
Similar to Table~\protect\ref{tab1} except for the time-odd NV coupling
constants inferred
from the time-even sector by using Eqs.~(\protect\ref{eq4.74a}) and
(\protect\ref{eq4.69a}) for a Fermi momentum of 1.35\,fm$^{-1}$.
}
\vspace*{1ex}
\begin{tabular}{|l|rr|}
\cline{2-3} \multicolumn{1}{l|}{}
         &  \multicolumn{1}{c}{$t=0$} &  \multicolumn{1}{c|}{$t=1$} \\
\hline
$C_t^{   s}      $ & $-$27.94756   &  $ $200.2052  \\
$C_t^{\Delta s}  $ & $ $20.92673   &  $ $23.21032  \\
$C_t^{   T}      $ & $-$26.47083   &  $-$44.78631  \\
\hline
\end{tabular}
\end{center}
\end{table}

The full density dependence of all Skyrme-force parameters was
recently implemented in a spherical self-consistent code
\cite{[Kor09]}. Here we use this implementation to test the results of
the NV expansion against the full-fledged solutions known for the Gogny
D1S force \cite{[Del10],[Del10a]}. In Table~\ref{tab4}, we show
results obtained for three sets of Skyrme-force parameters:
\begin{itemize}
\item S1Sa: fixed parameters
given in Table~\ref{tab2} for $k_F^0=1.35$\,fm$^{-1}$, used
together with the standard D1S parameters~\cite{[Ber91b]} of the zero-range spin-orbit
($W=130$\,MeV\,fm$^5$) and density-dependent
($t_3=1390.60$\,MeV\,fm$^4$) terms.
\item S1Sb: density-dependent parameters shown in Fig.~\ref{fig6},
implemented for $\rho=2k_F^3/3\pi^2$ and used with the standard values
of $W$ and $t_3$.
\item S1Sc: fixed parameters identical to those of S1Sa, but with the
value of $t_3=1385.35$\,MeV\,fm$^4$ slightly changed, so as to bring
the ground-state energy of $^{208}$Pb exactly to the true D1S value \cite{[Del10a]}.
\end{itemize}
Table~\ref{tab4} shows the ground-state energies $E$ of seven doubly
magic spherical nuclei, calculated by using the Skyrme-force parameters
S1Sa, S1Sb, and S1Sc, and compared with the Gogny-force energies
$E_G$. To facilitate the comparison, we also show relative
differences $\Delta{E}=(E-E_G)/|E_G|$ in percent along with the
corresponding RMS deviations (the last row in Table~\ref{tab4}).

On can see that for the Skyrme-force parameters S1Sa and S1Sb,
which are directly derived from the Gogny force by using the NV
expansion, one obtains the nuclear binding energies smaller by 1--2\%
as compared to those given by the original Gogny force. This should
be considered a very good result, although it cannot compete in
precision of describing experimental data with the original Gogny or
Skyrme forces, which have parameters directly fitted to experimental
binding energies. A simple rescaling of the parameter $t_3$ brings
the RMS deviation to below $0.4\%$, and makes the Skyrme force S1Sc
competitive with most other standard Skyrme parameterizations. At the
same time, the analogous RMS deviations obtained for the neutron and
proton radii are 0.20 and 0.30\% (S1Sa), 1.01 and 0.94\% (S1Sb),
and 0.26 and 0.44\% (S1Sc), respectively.

We note here in passing that in this study we defined the S1Sb
parameter set by considering the density-dependent coupling constants
$C_t^{\Delta\rho}$ (Fig.~\ref{fig6}) that multiply densities
$\rho_t\Delta\rho_t$. An attempt of using the same coupling constants
along with densities $-(\bbox{\nabla}\rho_t)^2$ gives, in fact, the
RMS deviations of binding energies (5.34\%), neutron radii (2.44\%),
and proton radii (2.27\%), which are significantly worse than those
of S1Sb (Table~\ref{tab4}). This shows that the prescription to
replace in EDFs integrated by parts (see
Refs.~\cite{[Dob96b],[Kor09]}) the Fermi momentum by
$\rho=2k_F^3/3\pi^2$ may lead to significantly different results.
Another recipy is to associate $k_F$ with density
before taking products of the two density matrices which means
both of the above terms will be active. However in order to
have a better correspondence with the traditional Skyrme functionals
we have made this association after taking the product.

It is not our purpose here to propose that any of the Skyrme-force
parameterizations introduced in the present work are better solutions
to the problem of finding the best agreement with data. It is already
known that within the standard second-order Skyrme-force
parameterizations, a spectroscopic-quality \cite{[Zal08]} force
cannot be found \cite{[Kor08]}. Nevertheless, it is gratifying to see
that the NV expansion allows us to bridge the gap between the
non-local and quasi-local EDFs, or between the finite-range and
zero-range effective forces. A more quantitative discussion of the
accuracy of the NV expansion will be possible by considering
higher-order NV expansions~\cite{[Car10]}.

\begin{table}
\begin{center}
\caption{\label{tab4}
Binding energies $E$ of seven doubly magic nuclei calculated
by using the Skyrme-force parameters
S1Sa, S1Sb, and S1Sc (see text) compared with the Gogny-force energies
$E_G$. All energies are in MeV.
}
\vspace*{1ex}
\begin{tabular}{|l|r|rr|rr|rr|}
\cline{2-8}  \multicolumn{1}{l|}{}
          &  \multicolumn{1}{c|}{D1S \protect\cite{[Del10a]}}
          &  \multicolumn{2}{c|}{S1Sa}
          &  \multicolumn{2}{c|}{S1Sb}
          &  \multicolumn{2}{c|}{S1Sc} \\
\cline{2-8}  \multicolumn{1}{l|}{}
          &  \multicolumn{1}{c|}{$E_G$}
          &  \multicolumn{1}{c}{$E$}  &  \multicolumn{1}{c|}{$\Delta{E}$}
          &  \multicolumn{1}{c}{$E$}  &  \multicolumn{1}{c|}{$\Delta{E}$}
          &  \multicolumn{1}{c}{$E$}  &  \multicolumn{1}{c|}{$\Delta{E}$} \\
\hline
$^{ 40}$Ca & $-$342.689    &  $-$335.312  & 2.15\% &  $-$340.642  & 0.60\% &  $-$339.369  &$ $0.97\% \\
$^{ 48}$Ca & $-$414.330    &  $-$409.118  & 1.26\% &  $-$410.698  & 0.88\% &  $-$414.213  &$ $0.03\% \\
$^{ 56}$Ni & $-$481.111    &  $-$473.497  & 1.58\% &  $-$471.970  & 1.90\% &  $-$479.843  &$ $0.26\% \\
$^{ 78}$Ni & $-$637.845    &  $-$630.447  & 1.16\% &  $-$629.066  & 1.38\% &  $-$638.837  &$-$0.16\% \\
$^{100}$Sn & $-$828.024    &  $-$814.568  & 1.63\% &  $-$814.896  & 1.59\% &  $-$826.453  &$ $0.19\% \\
$^{132}$Sn & $-$1101.670   &  $-$1086.272 & 1.40\% &  $-$1086.867 & 1.34\% &  $-$1101.445 &$ $0.02\% \\
$^{208}$Pb & $-$1637.291   &  $-$1612.634 & 1.51\% &  $-$1617.419 & 1.21\% &  $-$1637.291 &$ $0.00\% \\
\hline
RMS        &    n.a.       &     n.a.     & 1.56\% &     n.a.     & 1.33\% &     n.a.     &   0.39\% \\
\hline
\end{tabular}
\end{center}
\end{table}

Finally, in Fig.~\ref{fig5} we compare the time-odd coupling
constants calculated by using Eqs.~(\ref{eq20}) and (\ref{eq21}) with
those corresponding to the Skyrme-force parameters; that is,
calculated by using Eqs.~(\ref{eq4.74a}) and (\ref{eq4.69a}).
Similarly, Table~\ref{tab3} lists the numerical values of the
Skyrme-force time-odd coupling constants. As one can see, differences
between both sets of the time-odd coupling constants, shown in
Fig.~\ref{fig5} with open and full symbols, are quite substantial.
These results illustrate the fact that the NV expansion of the Gogny
force leads to the Skyrme functional and not to the Skyrme force.

We conclude this section by noting that functions $\pi_i(r)$
approximated by Gaussians (see
Eqs.~(\protect\ref{eq121a})--(\protect\ref{eq121c}) and
Fig.~\ref{fig1}) lead to the coupling constants and Skyrme-force
parameters, which, when plotted in the scales of
Figs.~\ref{fig4}, \ref{fig6}, and \ref{fig5}, are indistinguishable from those
presented in these figures.

\bigskip

\section{Conclusions}\label{sec9}

In the present study, we derived a set of compact expressions giving
the local energy density corresponding to the Kohn-Sham potential
energy for an arbitrary local finite-range central, spin-orbit, and
tensor interactions. The method is based on the Negele-Vautherin
density matrix expansion augmented by the odd-power gradient terms fulfilling
the gauge-invariance condition. The coupling constants of the local
energy density depend on a set of moments of the interaction
conforming to the ideas of the effective theory. The expansion is
based on the separation of scales between the range of the force and
space characteristics of the one-body density matrix. It leads to a
representation of dynamical properties of the system in terms of a set
of numbers, whereby complicated short-range characteristics of
effective interactions remain unresolved.

We pointed out the fact that to correctly describe the exchange
properties of the functional, proper treatment of the density matrix
in the nonlocal direction is essential. This immediately leads to the
local energy density that does not correspond to an averaged
zero-range pseudopotential. Therefore, the Negele-Vautherin expansion
performed up to NLO leads to the Skyrme functional and not to the
Skyrme force. Within this formalism, the only way to define the
Skyrme force is to match it to the time-even properties of the
non-local functional, disregarding those pertaining to the time-odd
channel.

We applied the general NLO expressions to the case of the central
finite-range part of the Gogny interaction. It turns out that the
obtained coupling constants of the local Skyrme functional quite
strongly depend on the Fermi momentum or on the density.
Nevertheless, the equation of state obtained for fixed coupling
constants, calculated at the saturation point of
$k_F=1.35$\,fm$^{-1}$, fairly well reproduces the exact Gogny-force
result. On the other hand, partial density dependence, inferred from
the explicit dependence of the coupling constants on $k_F$, gives
very unsatisfactory results.

By solving the self-consistent equations with the Skyrme-force
parameters derived from the Gogny force, one obtains an excellent
agreement (up to 1-2\%) of binding energies and radii with those
corresponding to the true Gogny force. This shows that the ideas of
the effective theory, whereby the finite-range nuclear forces are
sufficiently short-range to be replaced by contact quasi-potentials,
are applicable to low-energy nuclear observables.

We also discussed properties of the one-body density matrix for the
spin and isospin polarized infinite nuclear matter. In this case, one
obtains the nonlocal densities with mixed spin-isospin channels. As a
result, the local energy density is not invariant but covariant with
respect to rotational and isospin symmetries; that is, it does not have
the form of the standard Skyrme functional. It means that the
standard Negele-Vautherin expansion can only be performed for
unpolarized densities.

\bigskip

Interesting comments by Thomas Duguet are gratefully acknowledged.
This work was supported by the Polish Ministry of Science and Higher
Education under Contract No.\ N~N~202~328234, by the Academy of
Finland and University of Jyv\"{a}skyl\"{a} within the FIDIPRO
program, and by the U.S.\ Department of Energy under Contract Nos.\
DE-FC02-09ER41583 (UNEDF SciDAC Collaboration) and DE-FG02-96ER40963
(University of Tennessee).

\appendix

\section{Spin and isospin polarized infinite nuclear matter}\label{appA}

Let us consider the infinite nuclear matter with spin and isospin
polarizations, which is described by the one-body Hamiltonian,
\be\label{eq:301}
\hat{H}= -\frac{\hbar^2}{2m}\Delta - \bbox{H}\cdot\bbox{\sigma}
                                   - \vec{\lambda}\circ\vec{\tau},
\ee
where $\bbox{\sigma}$ and $\bbox{H}$ are the space vectors of
spin Pauli matrices and spin-polarization Lagrange multipliers,
respectively, and $\vec{\tau}$ and $\vec{\lambda}$ are the isovectors
of the analogous isospin Pauli matrices and isospin-polarization
Lagrange multipliers, whereas the dot ``$\cdot$'' (circle ``$\circ$'') denotes the scalar
(isoscalar) product. Each eigenstate of Hamiltonian (\ref{eq:301})
is a Slater determinant that depends on the orientations of the
Lagrange multipliers $\bbox{H}$ and $\vec{\lambda}$ in the space
and isospace, respectively. However, since the kinetic energy
is scalar and isoscalar, we can arbitrarily fix these orientations
to $\bbox{H}_z$ and $\vec{\lambda}_3$, which gives the nonlocal densities
for spin-up and spin-down ($\sigma=\pm1$) neutrons and protons ($\tau=\pm1$)
in the form,
\be\label{eq:303}
\bar{\rho}_{\sigma\tau}(\bboxr_1,\bboxr_2) =
\hint_{|\bbox{k}|<k_{F,\sigma\tau}} \rmd^3\bbox{k}\,
e^{i\bbox{k}\cdot(\bboxr_1-\bboxr_2)}
=\bar{\rho}_{\sigma\tau}\bar{\pi}_{\sigma\tau}(\bboxr) .
\ee
The system simply separates into four independent Fermi spheres
for spin-up and spin-down neutrons and protons,
with four constant densities $\bar{\rho}_{\sigma\tau}$,
whereas the dependence on the relative position vector $\bboxr$
is given by four scalar functions $\bar{\pi}_{\sigma\tau}(r)$
[compare Eq.~(\ref{eq13})],
\be\label{eq:305}
\bar{\pi}_{\sigma\tau}(r) = \frac{3j_1(k_{F,\sigma\tau}r)}{k_{F,\sigma\tau}r} .
\ee
We note that the spin-isospin indices $\sigma\tau$ pertain here to the
preselected quantization axis defined by the chosen directions
of $\bbox{H}$ and $\vec{\lambda}$, which define an ``intrinsic''
reference frame. In this reference frame, densities are marked with a
bar symbol.

The ground state of the system is obtained by filling the four
Fermi spheres up to the common Fermi energy $\epsilon_F$,
\be\label{eq:306}
\epsilon_F = \epsilon_{F,\sigma\tau} = \frac{\hbar^2k_{F,\sigma\tau}^2}{2m}
                                     - \bbox{H}_z\sigma
                                     - \vec{\lambda}_3\tau,
\ee
which defines the four Fermi momenta $k_{F,\sigma\tau}$.
Finally, by varying $\epsilon_F$, one obtains systems with different
total densities $\rho=\sum_{\sigma\tau}\bar{\rho}_{\sigma\tau}$.

It is, of course, clear that for the asymmetric and polarized infinite
nuclear matter, the density matrix of Eq.~(\ref{eq:302}) is diagonal
in spin and isospin,
\be\label{eq:307}
\bar{\rho}(\bboxr_1\sigma_1\tau_1,\bboxr_2\sigma_2\tau_2)
= \bar{\rho}_{\sigma_1\tau_1}(\bboxr_1,\bboxr_2)
\delta_{\sigma_1\sigma_2}
\delta_{\tau_1\tau_2}
\ee
and thus the nonlocal densities $\bar{\rho}_{\mu k}(\bboxr_1,\bboxr_2)$
have non-zero components only for $\mu=0$ or $z$ and $k=0$ or 3, that is,
\be\label{eq:308}
\left(\ba{c}
\bar{\rho}_{00}(\bboxr_1,\bboxr_2) \\
\bar{\rho}_{03}(\bboxr_1,\bboxr_2) \\
\bar{\rho}_{z0}(\bboxr_1,\bboxr_2) \\
\bar{\rho}_{z3}(\bboxr_1,\bboxr_2) \ea\right)
=
\left(\ba{rrrr}
   1 &    1 &    1 &    1  \\
   1 &   -1 &    1 &   -1  \\
   1 &    1 &   -1 &   -1  \\
   1 &   -1 &   -1 &    1  \ea\right)
\left(\ba{r@{}l}
\bar{\rho}_{++}&(\bboxr_1,\bboxr_2) \\
\bar{\rho}_{+-}&(\bboxr_1,\bboxr_2) \\
\bar{\rho}_{-+}&(\bboxr_1,\bboxr_2) \\
\bar{\rho}_{--}&(\bboxr_1,\bboxr_2) \ea\right) ,
\ee
where we have abbreviated the indices of $\sigma\tau$ just to their signs.
After expressing the right-hand side of this equation in terms
nonlocal densities (\ref{eq:303}), one obtains:
\be\label{eq:309}
\left(\ba{c}
\bar{\rho}_{00}(\bboxr_1,\bboxr_2) \\
\bar{\rho}_{03}(\bboxr_1,\bboxr_2) \\
\bar{\rho}_{z0}(\bboxr_1,\bboxr_2) \\
\bar{\rho}_{z3}(\bboxr_1,\bboxr_2) \ea\right)
=
\left(\ba{rrrr}
   \bar{\pi}_{00}(r) &   \bar{\pi}_{03}(r) & \bar{\pi}_{z0}(r)   & \bar{\pi}_{z3}(r)    \\
   \bar{\pi}_{03}(r) &   \bar{\pi}_{00}(r) & \bar{\pi}_{z3}(r)   & \bar{\pi}_{z0}(r)    \\
   \bar{\pi}_{z0}(r) &   \bar{\pi}_{z3}(r) & \bar{\pi}_{00}(r)   & \bar{\pi}_{03}(r)    \\
   \bar{\pi}_{z3}(r) &   \bar{\pi}_{z0}(r) & \bar{\pi}_{03}(r)   & \bar{\pi}_{00}(r)    \ea\right)
\left(\ba{l}
\bar{\rho}_{00} \\
\bar{\rho}_{03} \\
\bar{\rho}_{z0} \\
\bar{\rho}_{z3} \ea\right) ,
\ee
where functions $\bar{\pi}_{\mu k}(r)$ are defined similarly as in Eq.~(\ref{eq:308}), namely,
\be\label{eq:310}
\left(\ba{c}
\bar{\pi}_{00}(r) \\
\bar{\pi}_{03}(r) \\
\bar{\pi}_{z0}(r) \\
\bar{\pi}_{z3}(r) \ea\right)
= \frac{1}{4}
\left(\ba{rrrr}
   1 &    1 &    1 &    1  \\
   1 &   -1 &    1 &   -1  \\
   1 &    1 &   -1 &   -1  \\
   1 &   -1 &   -1 &    1  \ea\right)
\left(\ba{r@{}l}
\bar{\pi}_{++}&(r) \\
\bar{\pi}_{+-}&(r) \\
\bar{\pi}_{-+}&(r) \\
\bar{\pi}_{--}&(r) \ea\right) .
\ee
Already here we see the main problem: for the spin and isospin polarized systems,
the spin-isospin channels of nonlocal
densities $\bar{\rho}_{\mu k}(\bboxr_1,\bboxr_2)$ in Eq.~(\ref{eq:309})
are linear combinations
of the spin-isospin channels of local
densities $\bar{\rho}_{\mu k}$; that is, the spin-isospin channels become mixed.

To make the preceding result even more clear, we note that the
spin-isospin directions of the Lagrange multipliers $\bbox{H}$ and
$\vec{\lambda}$ can be arbitrarily varied and the spin-isospin
directions of the nonlocal densities $\rho_{\mu k}(\bboxr_1,\bboxr_2)$, local
densities $\rho_{\mu k}$, and functions  $\pi_{\mu k}(r)$ are always
aligned with those of the Lagrange multipliers. Therefore,
we can use the directions of the local densities instead of those
pertaining to the Lagrange multipliers. By using the standard
densities \cite{[Per04]} in the
(i)   scalar-isoscalar channel ($\rho=\rho_{00}$),
(ii)  vector-isoscalar channel ($\bbox{s}_\mu=\rho_{\mu 0}$, for $\mu=x,y,z$),
(iii) scalar-isovector channel ($\vec{\rho}_k=\rho_{0 k}$, for $k=1,2,3$), and
(iv)  vector-isovector channel ($\vec{\bbox{s}}_{\mu k}=\rho_{\mu k}$, for $\mu=x,y,z$ and $k=1,2,3$),
we then define functions $\pi(r)$ in the four channels as,
\be\label{eq:311}
\ba{rll@{}l}
\pi             (r) &=&                                        &\bar{\pi}_{00}(r), \\
\bbox{\pi}      (r) &=& \frac{\bbox{s}}{|\bbox{s}|}            &\bar{\pi}_{03}(r), \\
\vec{\pi}       (r) &=& \frac{\vec{\rho}}{|\vec{\rho}|}        &\bar{\pi}_{z0}(r), \\
\vec{\bbox{\pi}}(r) &=& \frac{\vec{\bbox{s}}}{|\vec{\bbox{s}}|}&\bar{\pi}_{z3}(r).
\ea\ee
Here, the ``intrinsic'' functions $\bar{\pi}_{\mu k}(r)$ do not depend
on the spin-isospin directions; that is, they are defined by
the following Fermi energies,
\be\label{eq:306a}
\epsilon_F = \epsilon_{F,\sigma\tau} = \frac{\hbar^2k_{F,\sigma\tau}^2}{2m}
                                     - |\bbox{H}|\sigma
                                     - |\vec{\lambda}|\tau.
\ee

Finally, definitions (\ref{eq:311}) allow us to present densities in
the ``laboratory'' reference frame as [compare Eq.~(\ref{eq:309})],
\bn
\label{eq:312a}
\rho          (\bboxr_1,\bboxr_2) &=& \rho          \,\pi(r) + \vec{\rho}    \circ  \vec{\pi}(r) + \bbox{s}      \cdot  \bbox{\pi}(r) + \vec{\bbox{s}}\cdot\circ\,\vec{\bbox{\pi}}(r)   , \\
\label{eq:312b}
\vec{\rho}    (\bboxr_1,\bboxr_2) &=& \vec{\rho}    \,\pi(r) + \rho               \,\vec{\pi}(r) + \vec{\bbox{s}}\cdot\,\bbox{\pi}(r) + \bbox{s}      \cdot       \vec{\bbox{\pi}}(r)   , \\
\label{eq:312c}
\bbox{s}      (\bboxr_1,\bboxr_2) &=& \bbox{s}      \,\pi(r) + \vec{\bbox{s}}\circ\,\vec{\pi}(r) + \rho               \,\bbox{\pi}(r) + \vec{\rho}         \circ  \vec{\bbox{\pi}}(r)   , \\
\label{eq:312d}
\vec{\bbox{s}}(\bboxr_1,\bboxr_2) &=& \vec{\bbox{s}}\,\pi(r) + \bbox{s}           \,\vec{\pi}(r) + \vec{\rho}         \,\bbox{\pi}(r) + \rho                    \,\vec{\bbox{\pi}}(r)   .
\en
Note that the same scalar-isoscalar function $\pi(r)$ multiplies
all local densities in the first terms of Eqs.~(\ref{eq:312a})--(\ref{eq:312d}).
Therefore, the postulate of using different functions in different
channels~\cite{[Geb09]} is not compatible with the results obtained
for the polarized nuclear matter.

Again we see that the spin-isospin channels of nonlocal densities are
mixed, namely, local densities in all channels contribute to every
channel in the nonlocal density. As a consequence, the energy density
is not invariant but only covariant with respect to the spin-isospin
rotations (see the discussion in the Appendix A of
Ref.~\cite{[Car08]}). Therefore, the NV expansion performed in the
polarized nuclear matter does not lead to the standard local
functional of Eq.~(\ref{eq18}). On the other hand, derivation in the
unpolarized nuclear matter corresponds to all functions
$\bar{\rho}_{\sigma\tau}(r)$ equal to one another, which leads to
vanishing functions $\vec{\pi}(r)$, $\bbox{\pi}(r)$, and
$\vec{\bbox{\pi}}(r)$. Then, in Eqs.~(\ref{eq:312a})--(\ref{eq:312d}),
only the first terms survive and the spin-isospin channels are not
mixed. Such a situation corresponds to postulating a
channel-independent function $\pi(r)$, which we employed in
Sec.~\ref{sec3}.

\bigskip

\bibliographystyle{unsrt}

\end{document}